\documentclass[lettersize,journal]{IEEEtran}
\usepackage{amsmath,amsfonts}
\usepackage{algorithm}
\usepackage{array}
\usepackage[caption=false,font=normalsize,labelfont=sf,textfont=sf]{subfig}
\usepackage{textcomp}
\usepackage{stfloats}
\usepackage{url}
\usepackage{verbatim}
\usepackage{graphicx}
\usepackage{cite}
\usepackage{cuted}
\usepackage{amsmath,amssymb,amsfonts}
\usepackage{graphicx}
\usepackage{textcomp}
\usepackage{soul}
\usepackage{setspace}
\usepackage{amsthm}
\usepackage{menukeys}
\usepackage{soul}
\usepackage{bbm}

\usepackage[compact]{titlesec}
\titlespacing{\section}{0pt}{2ex}{1ex}
\titlespacing{\subsection}{0pt}{2ex}{1ex}
\usepackage[table]{xcolor}
\definecolor{darkgreen}{rgb}{0,0.6,0}
\usepackage{tabularx}
\usepackage{nccmath}
\usepackage[strict]{changepage}

\usepackage{cite}
\usepackage[colorlinks=true, allcolors=blue]{hyperref}
\usepackage{tikz}
\usepackage{setspace}
\usetikzlibrary{arrows,shapes}
\usepackage{nicematrix}
\usepackage{arydshln}

\usepackage{float}           
\usepackage{algorithmicx}
\usepackage[noend]{algpseudocode}
\usepackage{arydshln}
\usepackage[colorinlistoftodos]{todonotes}

\makeatletter
\newcommand{\manuallabel}[2]{\phantomsection\def\@currentlabel{#2}\label{#1}}
\makeatother

\usepackage{titlesec}
\titlespacing*{\subsection}{0pt}{0.2\baselineskip}{0.2\baselineskip} 

\definecolor{der1}{RGB}{0,114,189}   
\definecolor{der2}{RGB}{217,83,25}   
\definecolor{der3}{RGB}{237,177,32}  
\definecolor{der4}{RGB}{126,47,142}  
\definecolor{der5}{RGB}{119,172,48}  

\newif\ifhighlightchanges
\highlightchangestrue       
\highlightchangesfalse    

\newcommand{\rev}[1]{%
    \ifhighlightchanges
        \hl{#1}%
    \else
        #1%
    \fi
}

\begin{document}

\title{Quantification and Regulation of Energy Reserves for Distributed Frequency and Voltage Control of Cyber-Physical Microgrids}

\author{Ahmed Saad Al-Karsani,~\IEEEmembership{Student Member,~IEEE,} Maryam Khanbaghi,~\IEEEmembership{Senior Member,~IEEE}
\thanks{Ahmed Saad Al-Karsani and Maryam Khanbaghi (corresponding author) are with Santa Clara University, 500 El Camino Real, Santa Clara, CA, USA. (email: \{ahsaad, mkhanbaghi\}@scu.edu)}
}

\markboth{ }%
{Shell \MakeLowercase{\textit{et al.}}: A Sample Article Using IEEEtran.cls for IEEE Journals}


\maketitle

\begin{abstract}
The introduction of Renewable Energy Sources (RES) and Distributed Energy Resources (DERs) has led to the formation of \rev{cyber-physical} Microgrids (MGs) and Networks of MGs (NMGs).  MGs and NMGs can operate in islanded mode, transforming the grid into a more distributed system. This has led to extensive studies in the literature on distributed hierarchical control strategies. Previous works have proposed distributed secondary level frequency and voltage regulation control schemes for Battery Energy Storage System (BESS)-based Grid-Forming (GFM) inverters with State of Charge (SoC) balancing. However, control applications in terms of service-based reserves and local resource adequacy in MGs are largely unexplored. {\color{black}Therefore, this paper proposes a BESS-based GFM inverter energy reserves framework, to quantify reserves for hierarchical control operation}. Additionally, to {\color{black}partially} regulate the proposed energy reserves, we propose the formulation of a {\color{black}modified} Distributed-Averaging Proportional-Integral (DAPI) controller with regulation energy reserve consensus. Modal analysis is performed to demonstrate the stability guarantees of energy consensus under simplified assumptions. Cyber-physical Controller Hardware-In-the-Loop (CHIL) simulation {\color{black}is performed on} an MG topologically based on the IEEE 13 bus test feeder system in MATLAB\textsuperscript{\textregistered}/Simulink\textsuperscript{\textregistered}. {\color{black}The proposed scheme results illustrate effective} frequency and voltage regulation {\color{black}along with improved power and energy sharing across droop-controlled and} Virtual Synchronous Machine (VSM) {\color{black}controlled} inverters.

\end{abstract}


\section{Introduction}\label{Section I}
The introduction of Renewable Energy Sources (RES) and Distributed Energy Resources (DERs) has brought forth a new school of thought for the structure of the power grid. It is postulated that the grid of the future would be an agglomeration of Microgrids (MGs) and Networks of MGs (NMGs), i.e., a collection of generation and load that resembles the grid at a smaller scale. As RES penetration increases at the distribution level, in turn, the power grid is shifting from its centralized and vertically integrated unidirectional generation to load structure to a more distributed one. Additionally, Inverter-Based Resources (IBRs), which differ in operation principles compared to conventional generation units, are increasingly adopted. Hence, new challenges are faced with respect to power system stability and operation.

In terms of operation, power systems have been conventionally governed by a hierarchical three-layer control. At the primary level, it is desired to achieve frequency and voltage stabilization as well as power sharing. Common primary level control schemes include droop control and Virtual Synchronous Machine (VSM)-control. Afterwards, secondary level control is applied to restore and maintain the frequency and voltage values to their nominal or reference values. Lastly, optimization of energy management with respect to technical or economic criteria is performed at the tertiary level.

While primary level control is mainly decentralized, secondary level control relies heavily on the communication infrastructure. Conventionally, Automatic Generation Control (AGC) has been utilized by Balancing Authorities (BAs) for frequency regulation. BAs send signals to regulation units to counteract the frequency deviation that results from imbalance between supply and demand in the grid. Accountability over BAs' regulation is usually governed by a net imbalance metric called Area Control Error (ACE). ACE is a measure of the imbalance resulting from differences between actual and scheduled interchanges, as well as a frequency bias term. One way of describing the disparity within interchanges is through the concept of inadvertent energy exchange, defined to account for such imbalances in power interchange, as well as the duration of imbalances \cite{b1}. {\color{black}The imbalances typically occur as a result of disturbances, i.e., load changes in the grid.}

{\color{black}In the context of MGs and NMGs, they are often more susceptible and sensitive to disturbances. This might require faster and localized secondary level control for regulation, instead of AGC.} In the literature, various distributed control schemes have been proposed at the secondary level per Nawaz \textit{et al.}'s review \cite{b2}. One prominent example is Distributed-Averaging Proportional-Integral (DAPI) control \cite{b3,b4}, a consensus control method for droop-controlled and VSM {\color{black}controlled} \cite{b5,b6} inverters. {\color{black}DAPI control uses} local interaction rules to achieve frequency and voltage regulation with power sharing. In terms of power losses due to frequency regulation, the transient performance study in \cite{b7} illustrated that {\color{black}including} frequency deviation within consensus dynamics acted as a damping factor. This has led to the superior DAPI performance over centralized control. {\color{black}However, losses accumulated from inadvertent energy exchanges are not explicitly addressed.}



\begin{figure}[!t]\vspace{-16pt}
  \centering

  \begin{minipage}[b]{0.48\textwidth}
    \centering
    \includegraphics[width=0.6\textwidth]{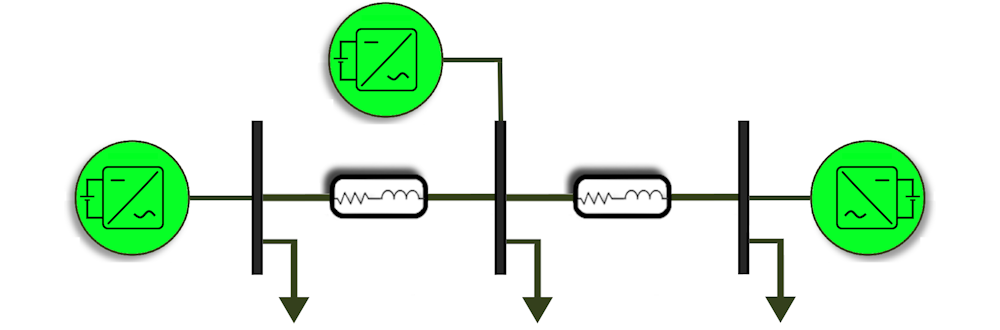}\\
    {\footnotesize (a)}
  \end{minipage}
  \begin{minipage}[b]{0.48\linewidth}
    \centering
    \includegraphics[width=\linewidth]{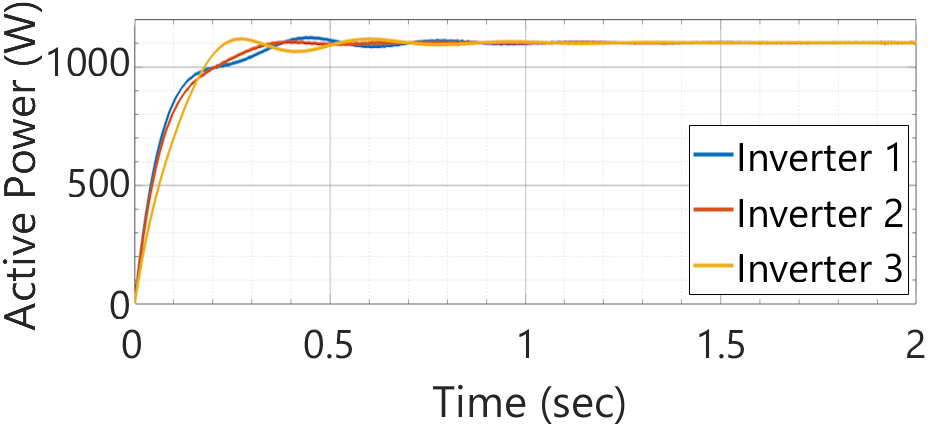}\\[-4pt]
    {\footnotesize (b)}
  \end{minipage}
  \hfill
  \begin{minipage}[b]{0.48\linewidth}
    \centering
    \includegraphics[width=\linewidth]{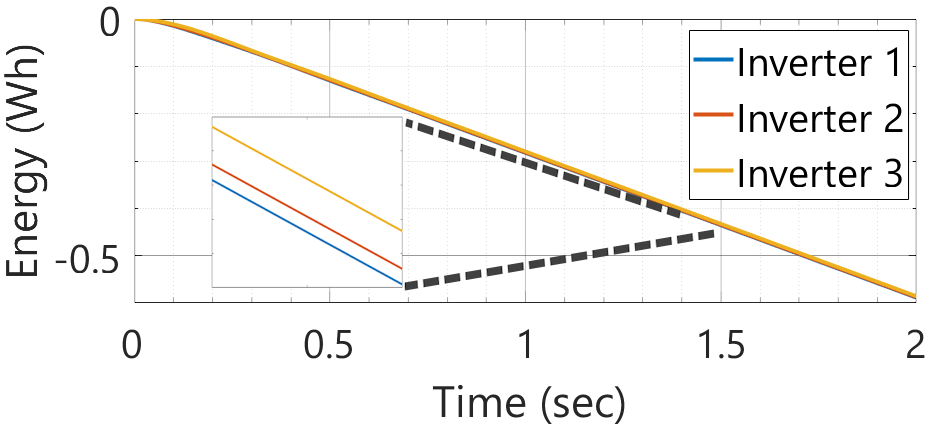}\\[-4pt]
    {\footnotesize (c)}\\
  \end{minipage}

  \caption{(a) Simulation model of three identical Virtual Synchronous Machine (VSM) {\color{black}controlled} inverters in MATLAB\textsuperscript{\textregistered}/Simulink\textsuperscript{\textregistered}, (b) active power output, (c) energy output. Power sharing is achieved with unequal energy discharge.}
  \label{Fig. 1}\vspace{-10pt}
\end{figure}

Nonetheless, distributed secondary level control in MGs and NMGs means that regulation likely comes at the cost of unequal (dis)charged energy from Battery Energy Storage Systems (BESS)-based Grid-Forming (GFM) inverters, analogous to inadvertent energy exchange. This is supported by Yan \textit{et al.} \cite{b8} in their review, who partly conclude that virtual inertia and frequency regulation have indirect energy-related costs to be considered. {\color{black}To better illustrate this problem, we  provide the} example of a simple MG with three identical VSM {\color{black}controlled} inverters. {\color{black}As shown in Fig. \ref{Fig. 1},} while active power sharing is achieved, inadvertent energy exchange occurs, which depends on topology, load and line characteristics. 

Thus, following primary level power sharing and secondary level regulation attempts, the inadvertent energy exchange results in energy expenditure from allocated reserves, which might later require compensatory reserve restoration efforts at the tertiary level \cite{b9}. This is based on the assumption that \textit{conventional generators have ample energy supply}, which is not necessarily the case with the small-scale and fast-varying RES and DERs such as BESS-based inverters in islanded grids. In essence, MG and NMG settings would benefit from accounting for energy sharing, i.e., equitable energy (dis)charging for BESS-based GFM inverters.



To perform energy sharing, there are various consensus-based State of Charge (SoC) sharing control schemes in the literature \cite{b10,b11,b12}, following the trend of SoC balancing with a focus on active power sharing and/or frequency regulation. Other prominent examples include \cite{b13,b14,b15,b16,b17,b18}. Such works establish distributed consensus control for primary and secondary level frequency and voltage regulation as well as SoC balancing. Further extensions of the works involve Controller Hardware-In-the-Loop (CHIL) validation \cite{b19,b20,b21} and robustness to communication noise and delays \cite{b22,b23,b24}. 

Generally, the SoC balancing works tend to deviate from the conventional hierarchical control structure. Fagundes \textit{et al.}'s review \cite{b25} on SoC balancing indicates that few relevant works link the balancing {\color{black}process with tertiary level energy management}. Tertiary level control typically concerns a service-centric framework, i.e., availability of headroom (reserves) for adequate operation. Per Rebours and Kirschen \cite{b26}, such a framework necessitates primary, secondary and tertiary level energy reserves for each generation unit. For BESS, examples include California Independent System Operator (CAISO) \cite{b27} and National Energy System Operator (NESO) \cite{b28} ancillary and market services {\color{black}imposing SoC constraints to obtain sufficient headroom for broader resource adequacy goals}. In spatially uneven MGs and NMGs, SoC balancing can adversely affect long-term \textit{local} headroom for resource adequacy. Additionally, reactive power sharing is not commonly addressed in such cases. Lastly, energy balancing in MG distribution level test cases with heterogeneous droop-based and VSM-based BESS inverters is not well explored. {\color{black}Therefore}, in this paper, we investigate the possibility of \textit{quantifying and regulating energy reserves for distributed frequency and voltage control of BESS-interfaced GFM inverters}. \rev{Unlike conventional DAPI control focusing on frequency and voltage regulation with power sharing, and SoC balancing schemes focusing on equalizing stored energy, the proposed scheme considers and regulates cumulative deviations from scheduled power setpoints.} The main contributions of this paper can be summarized as follows:
\begin{itemize}
    \item Quantification of the BESS-based GFM inverter energy reserves to define {\color{black}hierarchical} control headroom, based on the operating reserves concept in \cite{b26}.
    \item Regulation of primary and secondary level control reserves to {\color{black}implement distributed secondary frequency and voltage control with power and energy sharing.}
\end{itemize}
We perform modal analysis to demonstrate the stability guarantees of energy consensus under simplified assumptions. The proposed energy reserves regulation scheme is validated in a MATLAB\textsuperscript{\textregistered}/Simulink\textsuperscript{\textregistered} simulation of a distribution level MG based on the IEEE 13 bus distribution feeder system, including Arduino-based CHIL simulations. 

\begin{figure*}[t!]\manuallabel{Fig. 2}{2}\centering\vspace{-4pt}
  \includegraphics[width=0.9\textwidth]{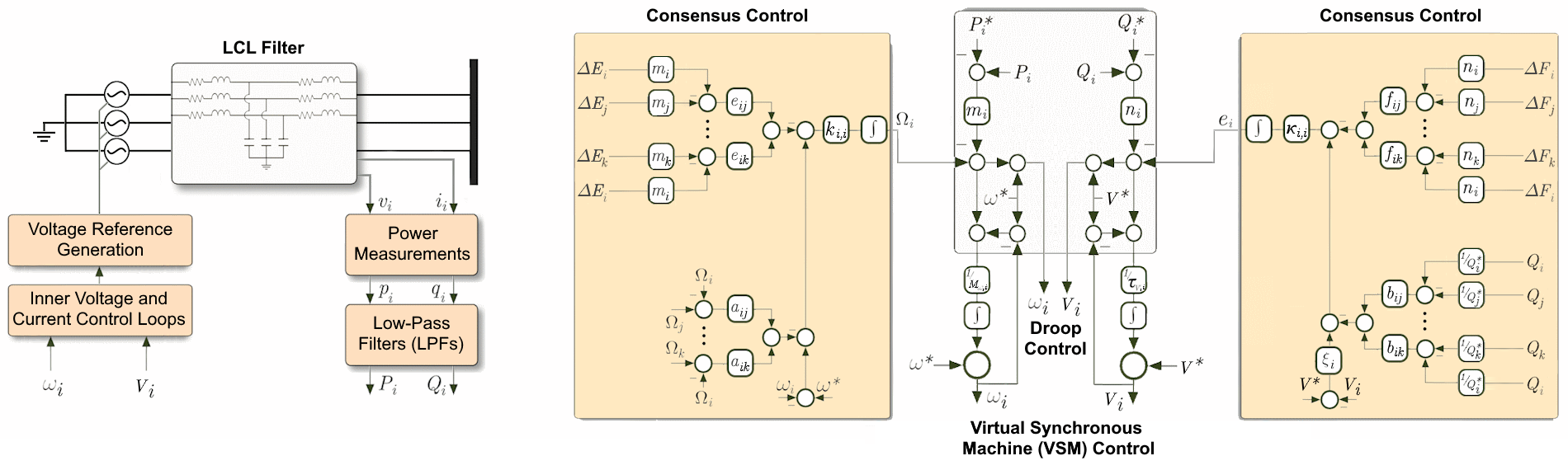}
  \caption{The average-based modeling of a Battery Energy Storage System (BESS)-based Grid-Forming (GFM) inverter with droop and Virtual Synchronous Machine (VSM) primary level control, Distributed-Averaging Proportional-Integral (DAPI) and the proposed energy consensus secondary level control. The instantaneous active and reactive powers $p_i,q_i$ are filtered using a Low-Pass Filter (LPF) to obtain the average power outputs.}
  \vspace{-9pt}
\end{figure*}

The rest of this paper is organized as follows: Section \ref{Section II} details the {\color{black}droop-based and} VSM-based GFM inverter with DAPI control, Section \ref{Section III} formulates the proposed energy reserves framework and control scheme, stability analysis is performed in Section \ref{Section IV}, simulation results are shown in Section \ref{Section V} before the concluding remarks in Section \ref{Section VI}.

\section{Grid-Forming (GFM) Inverter Modeling}\label{Section II}
In this study, an equivalent average-based model of the BESS-based GFM IBRs with small-scale storage is considered. We choose an average-based model over a switching-based model due to the latter's computational cost in simulations, and its negligible effects on primary and secondary level control. Both primary and secondary level control are performed in real-time with the assumption of no timescale separation except at the tertiary level. Fig. \ref{Fig. 2} illustrates the simulated model. Each inverter's average-based model control scheme consists of lower level, primary level and secondary level control and communication links with neighboring inverters as explained in the following subsections.

\subsection{Lower Level Control}
Per \cite{b29,b30}, to regulate the LC filter dynamics and account for $dq0$ axis coupling, we formulate the lower level inner current and outer voltage control loops. Since the test systems in Section \ref{Section V} are unbalanced in nature, we utilize lower level control in both positive-sequence and negative-sequence domains. 

For the $i^{th}$ inverter with outer voltage control loop, nodal analysis results in the following voltage dynamics \cite{b29,b30}:
\begin{subequations}\begin{align}
&\label{(1a)}C_f\dot{v}_{gd,i}(t) = i_{d,i}(t)-i_{gd,i}(t) + \omega_i(t) C_fv_{gq,i}(t) \\
&\label{(1b)}C_f\dot{v}_{gq,i}(t) = i_{q,i}(t)-i_{gq,i}(t) - \omega_i(t) C_fv_{gd,i}(t)
\end{align}\end{subequations}
where $C_f$ is the filter capacitance, $i_{d,i},i_{q,i}$ are the inverter-side $dq$ current, $v_{gd,i},v_{gq,i}$ and $i_{gd,i},i_{gq,i}$ are the grid-side $dq$ voltage and current respectively and $\omega_i$ is the angular frequency. Similarly, the inner current control loops rely on the following dynamics \cite{b29,b30}:
\begin{subequations}\begin{align}
&\label{(2a)}L_f\dot{i}_{d,i}(t) = v_{d,i}(t)-v_{gd,i}(t) + \omega_i(t) L_fi_{q,i}(t) - R_fi_{d,i}(t) \\
&\label{(2b)}L_f\dot{i}_{q,i}(t) = v_{q,i}(t)-v_{gq,i}(t) - \omega_i(t) L_fi_{d,i}(t) - R_fi_{q,i}(t)
\end{align}\end{subequations}
where $R_f,L_f$ are the filter resistance and inductance and $v_{d,i}(t) = V_i(t),v_{q,i}(t) = 0$ are the inverter-side $dq$ voltage, and $V_i(t)$ is the inverter {\color{black}primary level control} voltage. Accordingly, PI control is used to obtain the reference current $i_{dq,i}^*$ in the outer loop, before computing the $dq$ axis reference voltage $v_{dq,i}^*$ in the inner loop and eventually converted to $v_{abc,i}^*$. Details on gain calculations can be found in \cite{b29,b30}. Per \cite{b31}, {\color{black}to account for unbalanced operating conditions, we implement both positive-sequence and negative-sequence control using \eqref{(1a)}-\eqref{(2b)}. For negative-sequence control, $v_{d,i}(t) = 0$ and the sign of the inverter frequency $\omega_i$ is flipped.}  Lastly, the aforementioned equations rely on primary level frequency and voltage computation, detailed in the following subsection.

\subsection{Primary Level Control}
Following \cite{b32}, {\color{black}the conventional primary level droop control dynamics for a GFM inverter are defined as follows \cite{b32}:
\begin{subequations}\begin{align}
&\label{(3a)}\dot{\delta}_i(t) = \omega_i(t) - \omega^* \\
&\label{(3b)}\omega_i(t) = \omega^* - m_i(P_i(t) - P_i^*) \\
&\label{(3c)}V_i(t) = V^* - n_i(Q_i(t) - Q_i^*)
\end{align}\end{subequations}
where $\delta_i$ is the phase angle, $\omega_i$ and $\omega^*$ are the actual and setpoint frequency values, $m_i$ is the active power-frequency droop gain, $P_i$ and $P_i^*$ are the actual and setpoint active power outputs, $V_i$ and $V^*$ are the actual and setpoint voltage values, $n_i$ is the reactive power-voltage droop gain, $Q_i$ and $Q_i^*$ are the actual and setpoint reactive power outputs. In this work, $\omega^*$ and $V^*$ are fixed at $2\pi\cdot60$ rad/s and $1$ p.u.


Similarly,} the VSM-controlled inverter dynamics are defined as follows:
\begin{subequations}\begin{align}
&\label{(4a)}\dot{\delta}_i(t) = \omega_i(t) - \omega^* \\
&\label{(4b)}M_{\omega,i}\Delta\dot{\omega}_i(t) = -\Delta\omega_i(t) - m_i\Delta P_i(t) \\
&\label{(4c)}\tau_{V,i}\Delta\dot{V}_i(t) = -\Delta V_i(t) - n_i\Delta Q_i(t)
\end{align}\end{subequations}
where $\Delta \omega_i(t) = \omega_i(t) - \omega^*$, $\Delta P_i(t) = P_i(t) - P_i^*$, $\Delta V_i(t) = V_i(t) - V^*$, $\Delta Q_i(t) = Q_i(t) - Q_i^*$. Lastly, $M_{\omega,i}$ is the virtual inertia constant and $\tau_{V,i}$ is the voltage control time constant.

\manuallabel{Remark 1}{1}\textit{Remark 1.} The defined model in \eqref{(4a)}-\eqref{(4c)} resembles the generic primary level control derived by \cite{b32}. This means that the model can represent droop-derived dynamics. The notable differences between the droop-derived and VSM-derived dynamics are using instantaneous power instead of filtered (average) power, and the filter cutoff time constant instead of the respective VSM gains. We refer to \cite{b33} for more details.

Primary level control results in deviations in the form of steady-state errors following imbalances or disturbances. Thus, secondary level control is necessary for frequency and voltage regulation, as explained in the next subsection.

\subsection{Secondary Level Control}
The purpose of secondary level control is to eliminate or mitigate frequency and voltage deviations from reference values following primary level control action. To achieve this goal in a distributed fashion, we adopt Distributed-Averaging Proportional-Integral (DAPI) control \cite{b3,b4}. Accordingly, we add DAPI frequency $\Omega_i$ and voltage $\mathbf{e}_i$ secondary control terms to {\color{black}droop-controlled dynamics \eqref{(3b)} and \eqref{(3c)} as follows \cite{b33}:
\begin{equation}
\label{(5)}\Delta\omega_i(t) = - m_i\Delta P_i(t) + \Omega_i(t)
\end{equation}
\begin{equation}
\label{(6)}\Delta V_i(t) = - n_i\Delta Q_i(t) + \mathbf{e}_{i}(t)
\end{equation}
Additionally, the updated VSM dynamics in \eqref{(4b)} and \eqref{(4c)} with DAPI control are as follows:}
\begin{equation}
\label{(7)}M_{\omega,i}\Delta\dot{\omega}_i(t) = -\Delta\omega_i(t) - m_i\Delta P_i(t) + \Omega_i(t)
\end{equation}
\begin{equation}
\label{(8)}\tau_{V,i}\Delta\dot{V}_i(t) = -\Delta V_i(t) - n_i\Delta Q_i(t) + \mathbf{e}_{i}(t)
\end{equation}
such that the DAPI dynamics are defined as follows \cite{b3,b4}:
\begin{equation}\label{(6)} k_{i,i}\dot{\Omega}_i(t) = -\Delta \omega_i(t) - \sum_{j=1}^Na_{ij}(\Omega_i(t) - \Omega_j(t))\end{equation}
\begin{equation}\label{(7)}\kappa_{i,i}\dot{\mathbf{e}}_i(t) = -\beta_i\Delta V_i(t) - \sum_{j=1}^Nb_{ij}\left(\frac{Q_i(t)}{Q_i^*} - \frac{Q_j(t)}{Q_j^*}\right)\end{equation}
where $a_{ij},b_{ij}$ are the respective communication link gains, $k_{i,i},\kappa_{i,i}$ are the inverse integral frequency and voltage gains, $\beta_i$ is the voltage deviation gain and $N$ is the total number of cyber-physically linked inverters. \rev{If $Q^*_i$ or $Q^*_j$ are not taken to be the inverter rating, then $b_{ij}$ in} \eqref{(7)} \rev{is taken to be $0$ if $Q_i^*=0$ or $Q_j^*=0$.} It is known that there is a tradeoff between local voltage regulation and reactive power sharing, due to the voltage deviation associated with changing reactive power output, subject to topology and impedance.

As mentioned in Section \ref{Section I}, DAPI control can achieve frequency and voltage regulation along with power sharing. However, changes in load and/or network topology can result in uneven regulation energy (dis)charge, even across identical inverters. In the next section, we formulate an energy reserves framework to account for hierarchical control energy exchanges, along with a consensus-based energy reserves regulation scheme.

\section{Energy Reserve Formulation and Consensus}\label{Section III}

It is desired to establish real-time energy reserve metrics for each BESS-based inverter, accurate enough to consider inadvertent energy exchange resulting from RES and load variability. We can utilize the definition of hierarchical control reserves by Rebours and Kirschen \cite{b26}. Fig. \ref{Fig. 3} illustrates the allocated reserves for the different levels of conventional hierarchical control. 

\begin{figure}[b!]\manuallabel{Fig. 3}{3}\vspace{-8pt}
  \centering \includegraphics[width=0.72\columnwidth]{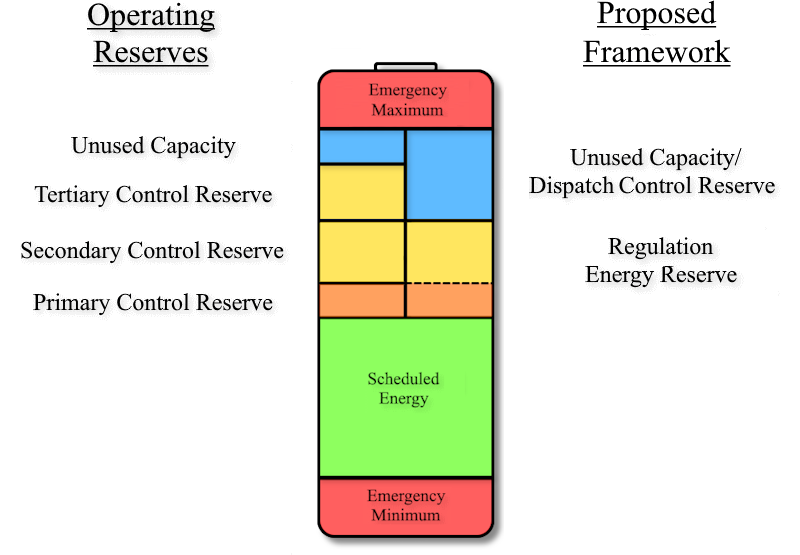}
    \caption{Allocation of Battery Energy Storage System (BESS) reserves for hierarchical control. Left: framework by Rebours and Kirschen \cite{b26}. Right: proposed framework for real-time primary and secondary level control and tertiary level control headroom.}
\end{figure}

It is worth noting that in the traditional power grid, there was a less urgent need for emphasizing and quantifying real-time inadvertent energy exchanges. This is in part due to the large-scale slow-varying nature of conventional generators, and fuel availability. Thus, the conventional approach introduces some redundancies with respect to the reserves. In this work, we propose that the reserves be redefined according to the deviation from power setpoint outputs.


\subsection{Energy Reserves Formulation}
If the inverter's power output deviates from its setpoint $P_i^*, Q_i^*$, then that will result in an energy supply and demand imbalance, i.e., inadvertent energy exchange. The energy imbalance accumulates at both transient and steady-state conditions until the setpoints are updated \rev{by an external} tertiary level \rev{control}. {\color{black}Accordingly, \rev{assuming $-P_{max,i} \leq P_i \leq P_{max,i}$ and $-Q_{max,i} \leq Q_i \leq Q_{max,i}$,} we propose to define each} inverter's energy reserves as a simple expression relating to the inadvertent energy exchange:
\begin{equation}\label{(8)}E_i(t) = \bar{E}_{c,i} - \Delta E_i(t) = \bar{E}_{c,i} - \int_{0}^{t} \Delta P_i(\tau)\ d\tau\end{equation}
\begin{equation}\label{(9)}F_i(t) = \bar{F}_{c,i}(t) - \Delta F_i(t) = \bar{F}_{c,i}(t) - \int_{0}^{t} \Delta Q_i(\tau)\ d\tau\end{equation}
where $E_i(t),F_i(t)$ are the active energy and reactive support unused (or dispatchable) capacity, $\bar{E}_{c,i}$ and $\bar{F}_{c,i}(t)$ are the total active and reactive support capacity and $\Delta E_i(t)$ and $\Delta F_i(t)$ are the active energy and reactive support imbalance, or inadvertent energy exchange. This framework is based on the assumption that any error between the actual and setpoint values (forecast ahead of time) would have to be compensated for using the defined regulation energy reserve metric.

\vspace{5.5pt}
\manuallabel{Remark 2}{2}\textit{Remark 2.} Unlike active energy, reactive energy (referred to as reactive support) does not \textit{directly} affect the BESS SoC. Therefore, we link it to the reactive power headroom instead. Nonetheless, reactive energy (in VARh) is used by power utility companies for power factor billing purposes \cite{b34}. Further discussion on related ancillary services and VARh pricing can be found in \cite{ancillarypaper}.
\vspace{5.5pt}

Since there is no direct energy measure equivalent for reactive power capacity, the reactive power capacity (headroom) can be computed using the inverter capacity $S_{max,i}$ \rev{(where $P_{max,i}^2 + Q_{max,i}^2 \leq S_{max,i}^2$)} as follows:
\begin{equation}\label{(10)}
    \bar{F}_{c,i}(t) = \int_{t-\Delta t}^{t} \max\left(\sqrt{S_{max,i}^2 - P_i(\tau)^2} - |Q_i(\tau)|,0\right)\ d\tau
\end{equation}
where $\Delta t$ is the time step and $S_{max,i}$ is the maximum apparent power based on either inverter or overcurrent limit, i.e., \rev{$S_{max,i}=min\{S_{rated,i}, V_i\bar{I}_i\}$} for inverter output voltage and maximum current $\bar{I}_i$. 

We note that in this work, our focus will be on achieving consensus for primary and secondary level regulation energy reserves of inverters. To achieve this goal, we can utilize the regulation energy imbalance expressions $\Delta E_i, \Delta F_i$ to eliminate inadvertent energy exchange across inverters. This can be done in a distributed manner using consensus-based control, as explained in the following subsection.

\subsection{Regulation Energy Reserve Consensus}
We aim to control the defined energy imbalances $\Delta E_i(t)$ and $\Delta F_i(t)$ in order to maintain equal regulation energy reserves across inverters. In other words, it is desired to {\color{black}achieve energy sharing by controlling regulation energy reserves}. This can be achieved within the DAPI control scheme, by introducing the following updated consensus terms {\color{black}into \eqref{(6)} and \eqref{(7)}}:
\begin{align}\label{(11)} k_{i,i}\dot{\Omega}_i(t) = & -\Delta \omega_i(t) - \sum_{j=1}^Na_{ij}(\Omega_i(t) - \Omega_j(t)) - \dots \notag\\ & \sum_{j=1}^Ne_{ij}(m_i\Delta E_i(t) - m_j\Delta E_j(t))\end{align}
\begin{align}\label{(12)}\kappa_{i,i}\dot{\mathbf{e}}_i(t) = & -\beta_i\Delta V_i(t) - \sum_{j=1}^Nb_{ij}\left(\frac{Q_i(t)}{Q_i^*} - \frac{Q_j(t)}{Q_j^*}\right) - \dots \notag\\ & \sum_{j=1}^Nf_{ij}\left(n_i\Delta F_i(t) - n_j\Delta F_j(t)\right)\end{align}
{\color{black}where $e_{ij}$ and $f_{ij}$ are the active and reactive regulation reserve communication link gains, respectively.} In \eqref{(11)} and \eqref{(12)}, if the newly added consensus terms go to zero, then every inverter's regulation energy reserve is at equal capacity for equal initial conditions. The droop gains are included as scalar multipliers to normalize the energy imbalances based on the respective inverter's capacity. We remark that consensus can indeed be achieved in the case of \eqref{(11)}. However, in the case of \eqref{(12)}, for nonzero $\beta_i$, there will be a tradeoff between voltage regulation, reactive power sharing and reactive support imbalance sharing. In the next section, we examine the stability of the proposed scheme for droop-controlled and VSM {\color{black}controlled} inverters.

\section{Small-Signal Stability Analysis}\label{Section IV}
In this section, we address the frequency stability and voltage stability for a composite MG {\color{black}model,} consisting of aggregated droop-controlled and/or VSM-controlled inverters as defined in \eqref{(3a)}-\eqref{(4c)}, \eqref{(11)} and \eqref{(12)}. For convenience of analysis, where $I$ is an $N\times N$ identity matrix, we assume homogeneous droop gains $M_P = mI, N_Q = nI$, inertia and voltage control constant values $M_\omega=m_\omega I$, $T_V=\tau_VI$, DAPI inverse integral gains $K=k_iI,\kappa=\kappa_i I$ and voltage deviation gains $B = \beta I$. {\color{black}In case of droop control, we can obtain an expression resembling VSM dynamics by either assuming $m_\omega = \tau_V = \varepsilon$ for a sufficiently small $\varepsilon$, or deriving droop-controlled dynamics with instantaneous power terms per \cite{b33}.} Additionally, the lossless line and $P-\omega$, $Q-V$ decoupling assumptions are utilized. Lastly, we assume there exists a synchronous {\color{black}steady-state} solution with sufficiently small phase angle differences and voltage deviations across buses.

\subsection{Frequency Stability}\label{Section IV-A}
We begin with the frequency DAPI control dynamics. {After \color{black}adopting \eqref{(4a)}, \eqref{(4b)}, \eqref{(11)} and $\Delta E_i$ and dropping the time notation, the composite system is represented as follows}:
\begin{subequations}\begin{align}
    &\label{(13a)}\dot{\delta} = \Delta \omega \\
    &\label{(13b)}m_\omega\Delta\dot{\omega} = -\Delta \omega -m\Delta P + \Omega \\
    &\label{(13c)}k_i\dot{\Omega} = -\Delta\omega - \mathcal{L}_a\Omega - m\mathcal{L}_e\Delta E \\
    &\label{(13d)}\Delta\dot{E} = \Delta P
\end{align}\end{subequations} 
for composite phase angle $\delta$, frequency deviation $\Delta \omega$, frequency consensus $\Omega$ and active energy imbalance $\Delta E$ terms. $\mathcal{L}_a$ and $\mathcal{L}_e$ represent the Laplacian matrices for frequency consensus and active energy imbalance communication links. We note that $\mathcal{L}_e = \gamma_e\mathcal{L}_a$, for a scalar multiplier $\gamma_e$. At steady state, an equilibrium is reached such that:
\begin{equation*}
    (\delta_e,\Delta \omega_e,\Omega_e,\Delta E_e){\color{black}^T} = (\bar{\delta}+\Delta\bar{\omega}t,\Delta\bar{\omega},\bar{\Omega},\Delta \bar{E}+c_Pt){\color{black}^T}
\end{equation*}
for a scalar $c_P$. Thus, {\color{black}both sides cancel out in \eqref{(13a)}} and \eqref{(13b)}-\eqref{(13d)} are updated as follows:
\begin{subequations}\begin{align}
    &\label{(14b)}0 = -\Delta\bar{\omega} -m\Delta\bar{P} + \bar{\Omega} \\
    &\label{(14c)}0 = -\Delta \bar{\omega} - \mathcal{L}_a\bar{\Omega} - {\color{black}m\gamma_e}\mathcal{L}_a\Delta \bar{E} \\
    &\label{(14d)}c_P = \Delta\bar{P}
\end{align}\end{subequations}
Pre-multiplying \eqref{(14c)} by $\mathbf{1}^T$ (which belongs to the left null space of $\mathcal{L}_a$ per \cite{b4,b35,b36}) results in the following:
\begin{equation*}
    \Delta \bar{\omega} = 0 \Leftrightarrow \bar{\omega} = \omega^*
\end{equation*}
This leads to the steady state consensus variable $\bar{\Omega} = cm$. Hence, the equilibrium point is updated as follows:
\begin{equation*}
    (\delta_e,\Delta \omega_e,\Omega_e,\Delta E_e){\color{black}^T} = (\bar{\delta},0,\bar{\Omega},\Delta \bar{E}+c_Pt){\color{black}^T}
\end{equation*}

We are interested in the small-signal stability of the aforementioned dynamics. Thus, by linearizing around the above equilibrium point, we obtain the following expression:
\begin{equation*}
    (\tilde{\delta},\Delta \tilde{\omega},\tilde{\Omega},\Delta \tilde{E}){\color{black}^T} = (\delta-\bar{\delta},\Delta\omega,\Omega-\bar{\Omega},\Delta E-(\Delta\bar{E}+c_Pt)){\color{black}^T}
\end{equation*}

For a lossless line with minimal voltage difference across buses, we can linearize the active power through Taylor series expansion to obtain an expression based on $\tilde{\delta}$ as follows \cite{b3}:
\begin{equation*}P \approx \bar{P} + \left.\frac{\partial P}{\partial \delta}\right|_{\bar{\delta}}(\delta-\bar{\delta}) = \bar{P} + L_P\tilde{\delta}
\end{equation*}
\begin{equation*}
    \Delta \dot{\tilde{E}} = \Delta\tilde{P} = \Delta P - \Delta\bar{P} = L_P\tilde{\delta}
\end{equation*}
where $L_P$ is the Jacobian. Under minimal phase angle differences across buses, we can assume $L_P$ is in Laplacian-like form. Hence, the updated system is defined as follows:
\begin{subequations}\begin{align}
    &\dot{\tilde{\delta}} = \Delta\tilde{\omega} \\
    &m_{\omega}\Delta\dot{\tilde{\omega}} = -\Delta\tilde{\omega}-mL_P\tilde{\delta} + \tilde{\Omega} \\
    &k_i\dot{\tilde{\Omega}} = -\Delta\tilde{\omega} - \mathcal{L}_a\tilde{\Omega} - m\gamma_e\mathcal{L}_a\Delta \tilde{E} \\
    &\Delta \dot{\tilde{E}} = L_P\tilde{\delta}
\end{align}\end{subequations}

We note that $\Delta \tilde{E}$ and $\tilde{\Omega}$ may converge to nonzero values within $span(\mathbf{1})$, which can complicate the process of proving convergence towards an equilibrium point. To address this, we can apply a linear transformation to decompose the states into agreement and disagreement terms. For $\Delta \tilde{E}$, the agreement terms reflect the power sharing dynamics, where every inverter converges towards the same value $m\Delta \tilde{P}$ (or $span(\mathbf{1})$) at steady state. More importantly, the disagreement term goes to zero at steady state, making it a point of interest in our stability analysis. Per \cite{b35,b36}, we consider the dynamics in the disagreement space orthogonal to that of the agreement space $span(\mathbf{1})$. This is done by constructing a projection matrix $R \in \mathbb{R}^{(N-1)\times N}$ such that \cite{b35,b36}:
\begin{equation}
    \begin{matrix}
        R\mathbf{1}=0, & RR^T = I_{N-1}, & R^TR = \Pi.
    \end{matrix}
\end{equation}
where $\Pi$ is the disagreement space transformation matrix $\Pi = I - \Psi = I - \frac{\mathbf{1}\mathbf{1}^T}{N}$ \cite{b35,b36}. 

Pre-multiplying the states by $R$ results in the disagreement space {\color{black}projected states} $(\tilde{\delta}_\perp,\Delta \tilde{\omega}_\perp,\tilde{\Omega}_\perp,\Delta \tilde{E}_\perp){\color{black}^T}= (R\tilde{\delta},R\Delta \tilde{\omega},R\tilde{\Omega},R\Delta \tilde{E}){\color{black}^T}$, such that, for minimal deviations, $(\tilde{\delta},\Delta \tilde{\omega},\tilde{\Omega},\Delta \tilde{E}){\color{black}^T} = (R^T\tilde{\delta}_\perp,R^T\Delta \tilde{\omega}_\perp,R^T\tilde{\Omega}_\perp,R^T\Delta \tilde{E}_\perp){\color{black}^T}$. Now, to perform modal analysis, we substitute $\dot{\tilde{x}}_\perp = \lambda \tilde{x}_\perp$ for the projected states $\tilde{x}_\perp$ as follows:
\begin{subequations}\begin{align}
    &\label{(17a)}\lambda\tilde{\delta}_\perp = \Delta\tilde{\omega}_\perp \\
    &\label{(17b)}m_{\omega}\lambda\Delta\tilde{\omega}_\perp = -\Delta\tilde{\omega}_\perp-mL_{P\perp}\tilde{\delta}_\perp + \tilde{\Omega}_\perp \\
    &\label{(17c)}k_i\lambda\tilde{\Omega}_\perp = -\Delta\tilde{\omega}_\perp - \mathcal{L}_{a\perp}\tilde{\Omega}_\perp - m\gamma_e\mathcal{L}_{a\perp}\Delta \tilde{E}_\perp \\
    &\label{(17d)}\lambda\Delta\tilde{E}_\perp = L_{P\perp}\tilde{\delta}_\perp
\end{align}\end{subequations}
where $L_{P\perp} = RL_PR^T$ and $\mathcal{L}_{a\perp} = R\mathcal{L}_aR^T$. If we solve for $\tilde{\delta}_\perp$, then substituting \eqref{(17a)} into \eqref{(17b)} then \eqref{(17b)} and \eqref{(17d)} into \eqref{(17c)} results in the following characteristic polynomial $D_\omega$:
\begin{align}
    D_\omega =\ & \alpha_0\lambda^4 + \alpha_1\lambda^3 + \alpha_2\lambda^2 + \alpha_3\lambda + \alpha_4 \notag \\ =\ & (k_im_\omega I)\lambda^4 + (k_iI+m_\omega\mathcal{L}_{a\perp})\lambda^3 + \dots \notag \\ & (\mathcal{L}_{a\perp}+m k_i L_{P\perp} + I)\lambda^2 + (m\mathcal{L}_{a\perp} L_{P\perp})\lambda + \dots \notag \\ & m\gamma_e\mathcal{L}_{a\perp} L_{P\perp} = 0
\end{align}

We wish to evaluate the characteristic polynomial using the Routh-Hurwitz criterion \cite{b4}. This requires scalar modal decomposition, which in turn requires $\mathcal{L}_{a\perp}$ and $L_{P\perp}$ to commute. In other words, the physical and cyber links are sufficiently similar and simultaneously diagonalizable. With this assumption, we can now apply the Routh-Hurwitz criterion, which requires the following conditions to be true:
\begin{align*}
    & \alpha_0,\alpha_1,\alpha_2,\alpha_3,\alpha_4 > 0 \\
    & \alpha_1\alpha_2 > \alpha_0\alpha_3 \\
    & \alpha_1\alpha_2 > \alpha_1^2\alpha_4\alpha_3^{-1} + \alpha_0\alpha_3
\end{align*}
The first condition is true as all expressed terms are positive (in the projected disagreement space). Next, for the condition $(k_iI+m_\omega\mathcal{L}_{a\perp})(\mathcal{L}_{a\perp}+mk_i L_{P\perp} + I) > k_im_\omega m \mathcal{L}_{a\perp} L_{P\perp}$, it is clear that the expanded left hand side contains the right hand side term plus positive terms. Lastly, $(k_iI+m_\omega\mathcal{L}_{a\perp})(\mathcal{L}_{a\perp}+mk_i L_{P\perp} + I) > (k_iI+m_\omega\mathcal{L}_{a\perp})(k_iI+m_\omega\mathcal{L}_{a\perp}) + (mk_im_\omega \gamma_e\mathcal{L}_{a\perp} L_{P\perp})$ expands to $k_i(I - \gamma_ek_i)I + (k_i + m_\omega - 2k_i\gamma_em_\omega)\mathcal{L}_{a\perp} + m_\omega(1-\gamma_em_\omega)\mathcal{L}_{a\perp}^2 + mk_i^2L_{P\perp} > 0$. The expanded condition is true for $0 \leq k_i \leq 
\frac{1}{\gamma_e}$ and $0 \leq m_\omega \leq \frac{1}{\gamma_e}$. {\color{black}Since the Routh-Hurwitz stability conditions are met, we can certify that the small-signal frequency stability is guaranteed for the composite system in \eqref{(17a)}-\eqref{(17d)} projected into the disagreement space, with bounded $m_\omega$ and $k_i$ values. }

\subsection{Voltage Stability}
For voltage stability, we follow Simpson-Porco \textit{et al.}'s \cite{b4} small-signal stability analysis. {\color{black}For simplification purposes,} the reactive power output $Q_i$ can be redefined {\color{black}as a voltage expression} as follows \cite{b4}:
\begin{equation}
    Q_i = -V_i^2Y_{ii} + V_i\sum_{j=1}^NY_{ij}(V_i-V_j)
\end{equation}
for Kron-reduced local bus admittance $Y_{ii}$ and neighboring bus link admittance  $Y_{ij}$. Now, the composite system {\color{black}defined by \eqref{(4c)}, \eqref{(12)} and $\Delta F_i$} is represented as follows \cite{b4}:
\begin{subequations}\begin{align}
    &\tau_V\Delta\dot{V} = -\Delta V -n\Delta Q + \mathbf{e} \\
    &\kappa_i\dot{\mathbf{e}} = -\beta\Delta V - Q^{*^{-1}}\mathcal{L}_b\Delta Q - n\mathcal{L}_f\Delta F \\
    &\Delta\dot{F} = \Delta Q \\
    &\Delta Q = [V]YV-Q^*
\end{align}\end{subequations}
for the composite terms of voltage deviation $\Delta V$, reactive power deviation $\Delta Q$, voltage consensus $\mathbf{e}$ and reactive support imbalance $\Delta F$, the Laplacian communication link matrices $\mathcal{L}_b$ and $\mathcal{L}_f = \gamma_f\mathcal{L}_b$ and the respective Kron-reduced admittance matrix $Y$. We replace $Q$ in \eqref{(12)} with $\Delta Q$ by adding $-1 \equiv \frac{Q_i^*}{Q_i^*}$ and $+1 \equiv \frac{Q_j^*}{Q_j^*}$.

Now, as stated in the previous subsection, small-signal stability is examined around the equilibrium point $(\Delta \bar{V}, \bar{\mathbf{e}}, \Delta \bar{F} + c_Qt){\color{black}^T}$ by examining $(\Delta \tilde{V} = \Delta V - \Delta \bar{V}, \tilde{\mathbf{e}} = \mathbf{e} - \bar{\mathbf{e}}, \Delta\tilde{F} = \Delta F - (\Delta \bar{F} + c_Qt)){\color{black}^T}$. For sufficient uniformity in impedance, inverter parameters and cyber-physical links where $\mathcal{L}_b$ and $L_Q$ are sufficiently similar and simultaneously diagonalizable, we can project the relevant states into the disagreement subspace $(\Delta \tilde{V}_\perp, \tilde{\mathbf{e}}_\perp, \Delta\tilde{F}_\perp){\color{black}^T} = (R\Delta \tilde{V}, R\tilde{\mathbf{e}}, R\Delta\tilde{F}){\color{black}^T}$ such that $(\Delta \tilde{V}, \tilde{\mathbf{e}}, \Delta\tilde{F}){\color{black}^T} = (R^T\Delta \tilde{V}_\perp, R^T\tilde{\mathbf{e}}_\perp, R^T\Delta\tilde{F}_\perp){\color{black}^T}$. This leads to the newly formed state-space model as follows:
\begin{subequations}\begin{align}
    &\tau_V\Delta\dot{\tilde{V}}_\perp = -\Delta \tilde{V}_\perp -n\Delta \tilde{Q}_\perp + \tilde{\mathbf{e}}_\perp \\
    &\kappa_i\dot{\tilde{\mathbf{e}}}_\perp = -\beta\Delta \tilde{V}_\perp - Q^{*^{-1}}\mathcal{L}_{b\perp}\Delta \tilde{Q}_\perp - n\gamma_f\mathcal{L}_{b\perp}\Delta \tilde{F}_\perp \\
    &\Delta\dot{\tilde{F}}_\perp = \Delta \tilde{Q}_\perp
\end{align}\end{subequations}
where $\Delta\tilde{Q}_\perp \approx L_{Q\perp}\Delta\tilde{V}_\perp = R([\bar{V}]Y + [Y\bar{V}])R^T\Delta\tilde{V}_\perp$ and $[Q^*]^{-1}=Q^{*^{-1}}I$ for a nonzero $Q^{*}$. Now, we can perform modal analysis with $\dot{\tilde{x}}_\perp = \lambda \tilde{x}_\perp$ as follows:
\begin{subequations}\begin{align}
    &\label{(22a)}\tau_V\lambda\Delta\tilde{V}_\perp = -(I + nL_{Q\perp})\Delta\tilde{V}_\perp + \tilde{\mathbf{e}}_\perp \\
    &\label{(22b)}\kappa_i\lambda\tilde{\mathbf{e}}_\perp = -(\beta + Q^{*^{-1}}\mathcal{L}_{b\perp} L_{Q\perp})\Delta\tilde{V}_\perp - n\gamma_f\mathcal{L}_{b\perp}\Delta \tilde{F}_\perp \\
    &\label{(22c)}\lambda\Delta\tilde{F}_\perp = L_{Q\perp}\Delta\tilde{V}_\perp
\end{align}\end{subequations}
Substituting \eqref{(22a)} and \eqref{(22c)} into \eqref{(22b)} results in the following characteristic polynomial $D_V$:
\begin{align}
    D_V =\ & \alpha_0\lambda^3 + \alpha_1\lambda^2 + \alpha_2\lambda + \alpha_3 \notag\\ =\ & (\kappa_i \tau_V I)\lambda^3 + \kappa_i(I + nL_{Q\perp})\lambda^2 + \dots \notag\\ & (\beta + Q^{*^{-1}}\mathcal{L}_{b\perp} L_{Q\perp})\lambda + n\gamma_f\mathcal{L}_{b\perp} L_{Q\perp} = 0
\end{align}

Per \cite{b4} and following Section \ref{Section IV-A}, we can utilize the Routh-Hurwitz criterion to determine stability, which dictates the following conditions:
\begin{subequations}\begin{align}
    &\alpha_0,\alpha_1,\alpha_2,\alpha_3 > 0 \\
    & \alpha_1\alpha_2 > \alpha_0\alpha_3
\end{align}\end{subequations}
For the first condition, it is clear that it is true as all expressed terms are positive. The second condition is expressed as $(I + nL_{Q\perp})(\beta + Q^{*^{-1}}\mathcal{L}_{b\perp} L_{Q\perp}) > n\tau_V\gamma_f\mathcal{L}_{b\perp} L_{Q\perp}$. It is true if $Q^{*^{-1}} \geq n\tau_V\gamma_f$, and $n$ is typically related to the inverse maximum $Q$ deviation which is less than or equal to $Q^{*^{-1}}$ and/or {\color{black}$\tau_V,\gamma_f$ are sufficiently small. Since the Routh-Hurwitz stability conditions are met, we conclude that small-signal voltage stability in the projected disagreement space is guaranteed for the proposed composite system consisting of \eqref{(22a)}-\eqref{(22c)}.}

In the following section, we outline our simulation results to validate the proposed energy reserves consensus scheme.

\section{Simulation Results}\label{Section V}
To test the proposed scheme, an MG (Fig. \ref{Fig. 4}) that is topologically based on an islanded IEEE 13 bus test feeder system has been simulated in MATLAB\textsuperscript{\textregistered}/Simulink\textsuperscript{\textregistered}. In the model, there are three BESS-based GFM inverters. Table \ref{Table I} illustrates the relevant parameters. 

\begin{figure}[h]\manuallabel{Fig. 4}{4}
    \centering

    \begin{minipage}{0.48\columnwidth}
        \centering
        \includegraphics[width=0.9\linewidth]{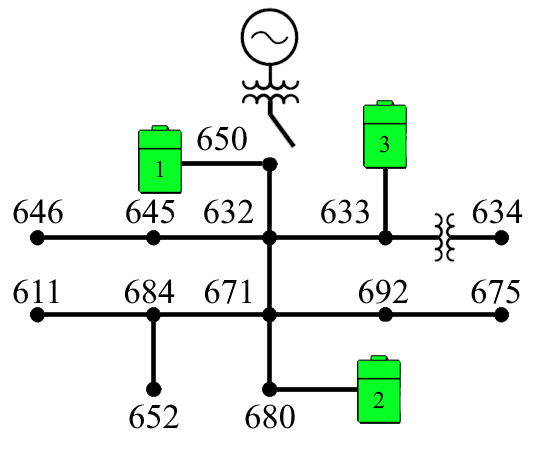}

        \vspace{0.25em}
        \footnotesize{(a)}
    \end{minipage}
    \hfill
    \begin{minipage}{0.48\columnwidth}
        \centering
        \includegraphics[width=\linewidth]{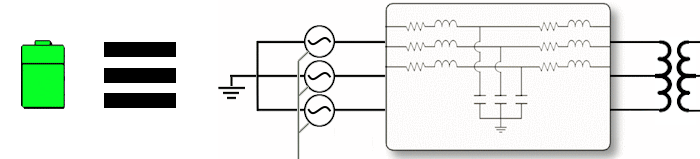}

        \vspace{0.25em}
        \footnotesize{(b)}
    \end{minipage}
    \vspace{4pt}
    \caption{(a) The Microgrid (MG) topologically based on an islanded IEEE 13 bus test feeder system, with three Battery Energy Storage System (BESS)-based inverters. (b) The BESS-based inverter average-based model.}\vspace{-8pt}
\end{figure}

\begin{table}[t!]\manuallabel{Table I}{I}\smaller\vspace{-10pt}
\caption{BESS-Based Grid-Forming (GFM) Inverter Parameters}
\centering
\begin{tabular}{c|c|c}
\hline
\hline
Parameter & Symbol & Value  \\
\hline
\hline
Inverter Rating & $S_{max,i}$ & 2.5 MVA \\
Nominal Frequency & $\omega^*$ & 2$\pi\cdot$60 rad/s\\
Nominal AC Voltage & $V^*$ & 480$\frac{\sqrt{2}}{\sqrt{3}}$ V\\
DC Voltage & $V_{dc,i}$ & 960$\frac{\sqrt{2}}{\sqrt{3}}$ V\\
Setpoint Active Power & $P^*_i$ & 1.2 MW \\
Setpoint Reactive Power & $Q^*_i$ & 0.6 MVAR \\
P-$\omega$ Droop Gain & $m_i$ & $2\cdot10^{-6}$ $\frac{rad}{W\cdot s}$\\
Q-V Droop Gain & $n_i$ & $2\cdot 10^{-5}$ $\frac{V}{VAR}$ \\
Consensus Voltage Deviation Term Gain  & $\beta_{i}$ & 0.1 \\
Inverse Integral Frequency Gain  & $k_{i,i}$ & 0.05 sec \\
Inverse Integral Voltage Gain  & $\kappa_{i,i}$ & 0.05 sec \\
Frequency Consensus Gain & $a_{ij}$ & 1 \\
Voltage Consensus Gain & $b_{ij}$ & 1 \\
Active Energy Reserve Consensus Gain & $e_{ij}$ & 0.5 \\
Reactive Support Reserve Consensus Gain & $f_{ij}$ & 0.05 \\
\hline
\hline
\end{tabular}
\vspace{-16pt}
\end{table}\normalsize

Additionally, we perform real-time cyber-physical CHIL validation using an OPAL-RT OP4510 test bench and Arduino microcontrollers, as shown in Fig. \ref{Fig. 5}. Each Arduino has a dedicated IP address and is connected to an Ethernet switch for communication with Simulink and neighboring Arduino units. The Arduino units receive the active and reactive power values $P_i,Q_i$ from the Simulink model. Additionally, they exchange distributed control variables, namely $\Omega_j, Q_j, \Delta E_j$ and $\Delta F_j$. In this simulation, each inverter uses DAPI control, with the setpoints $P^*$ and $Q^*$ determined based on total MG load. Three load pickup and load change scenarios are performed, based on separate active and reactive reserve consensus studies. The load changes occur at bus 675. In the figures, the reserve consensus is presented as the differences between inverter 1 and inverter 2 in blue, and inverter 2 and inverter 3 in red. 

\begin{figure}[t!]\manuallabel{Fig. 5}{5}
    \vspace{-6pt}
    \centering
    \includegraphics[width=\columnwidth]{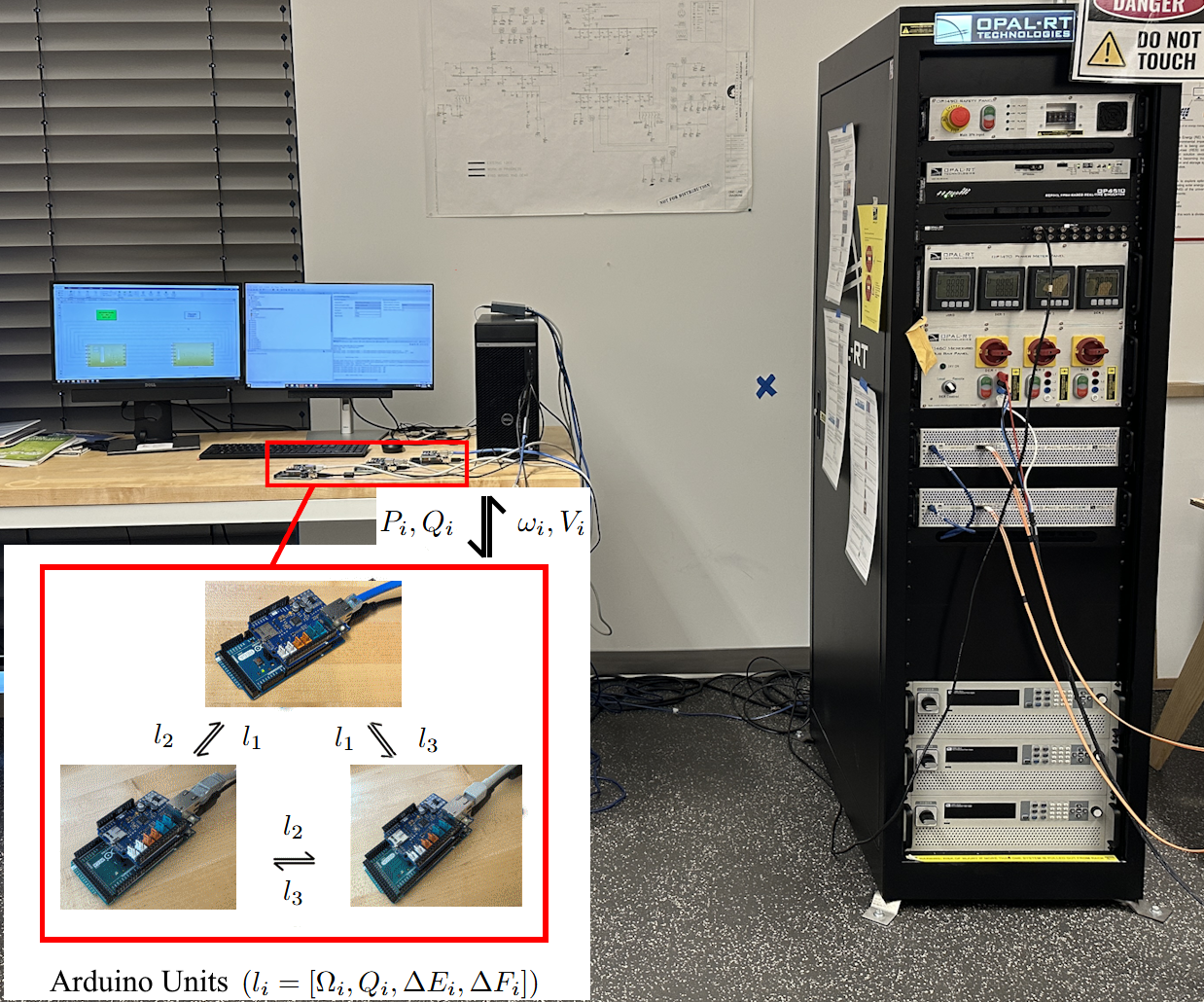}
    \caption{The Controller Hardware-In-the-Loop (CHIL) setup with OPAL-RT OP4510 and Arduino microcontrollers.}\vspace{-12pt}
\end{figure}

\begin{figure*}\vspace{-8pt}
  \centering
  \begin{minipage}[t]{0.48\textwidth}
    \centering
    \includegraphics[width=0.9\textwidth]{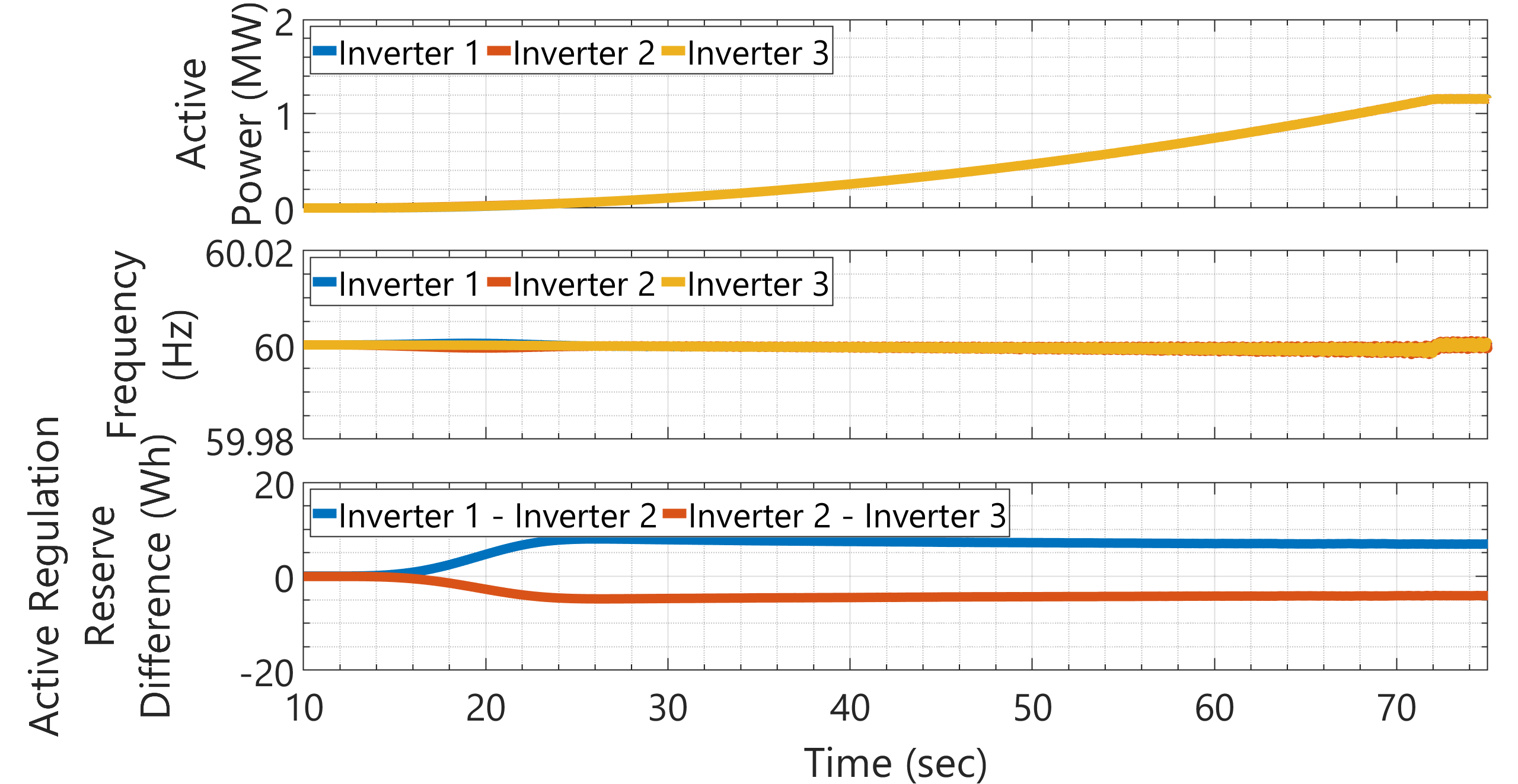}\vspace{-2pt}
    \caption{%
      From top to bottom: active power, frequency, and active regulation energy reserves using the base secondary control scheme from 10 to 75 seconds (load pickup for droop-controlled inverters).
    }
    \label{Fig. 6}
  \end{minipage}%
  \hfill
  \begin{minipage}[t]{0.48\textwidth}
    \centering
    \includegraphics[width=0.9\textwidth]{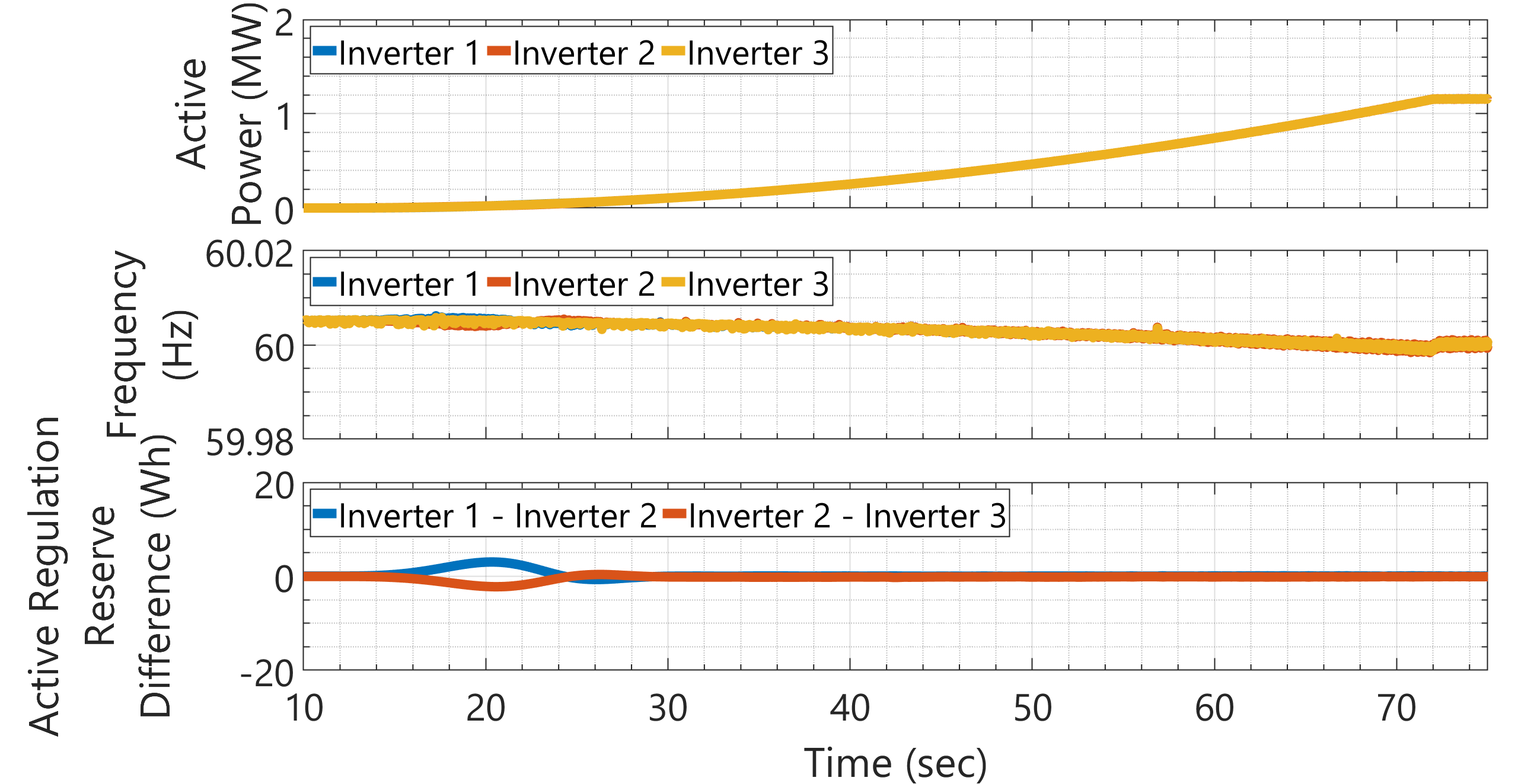}\vspace{-2pt}
    \caption{%
      From top to bottom: active power, frequency, and active regulation energy reserves using the proposed energy consensus scheme from 10 to 75 seconds (load pickup for droop-controlled inverters).
    }
    \label{Fig. 7}
  \end{minipage}
\end{figure*}

\begin{figure*}\vspace{-8pt}
  \begin{minipage}[t]{0.48\textwidth}
    \centering
    \includegraphics[width=0.9\textwidth]{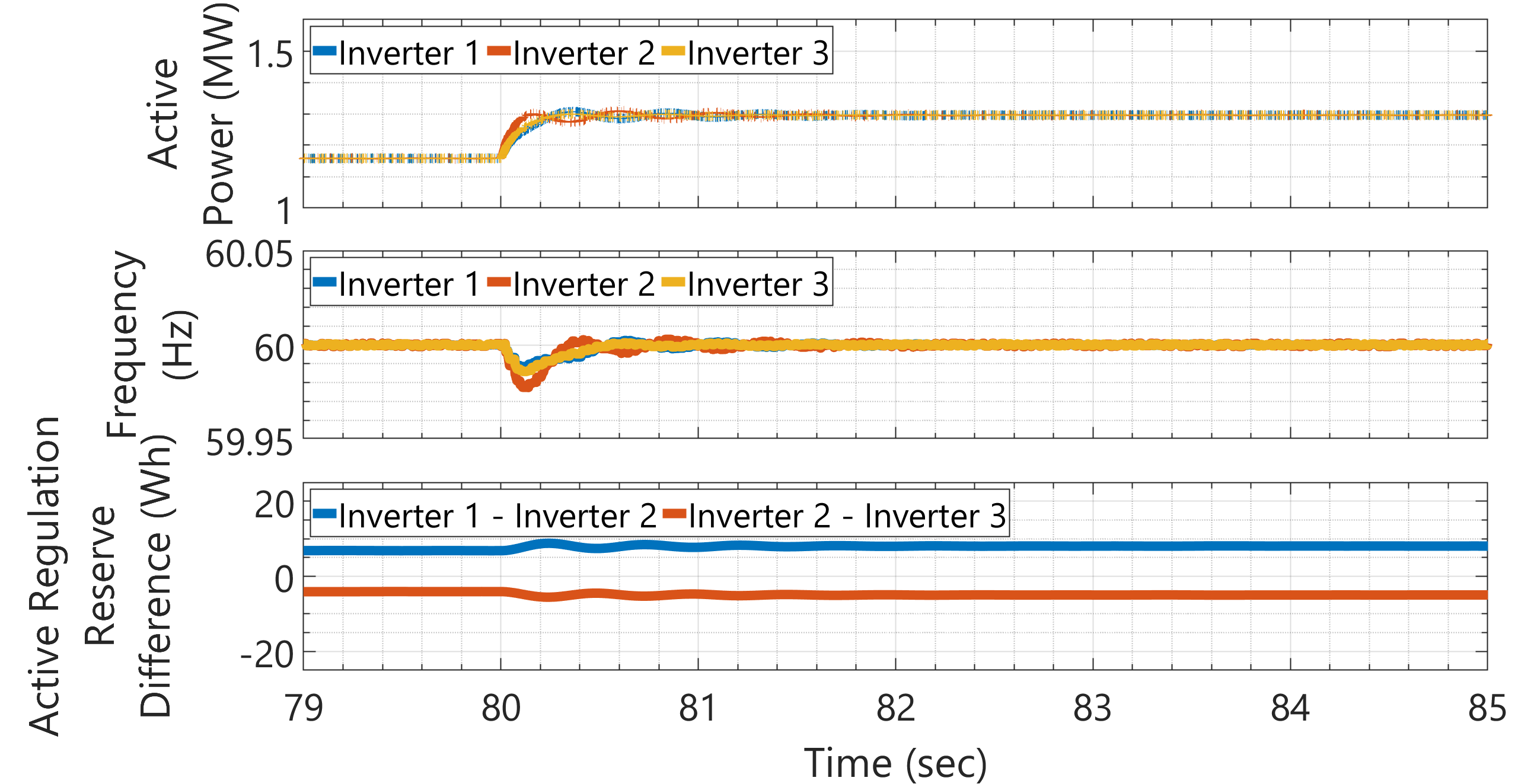}\vspace{-2pt} 
    \caption{%
      From top to bottom: active power, frequency, and active regulation energy reserves using the base secondary control scheme (load increase at 80 seconds for droop-controlled inverters).
    }
    \label{Fig. 8}
  \end{minipage}%
  \hfill
  \begin{minipage}[t]{0.48\textwidth}
    \centering
    \includegraphics[width=0.9\textwidth]{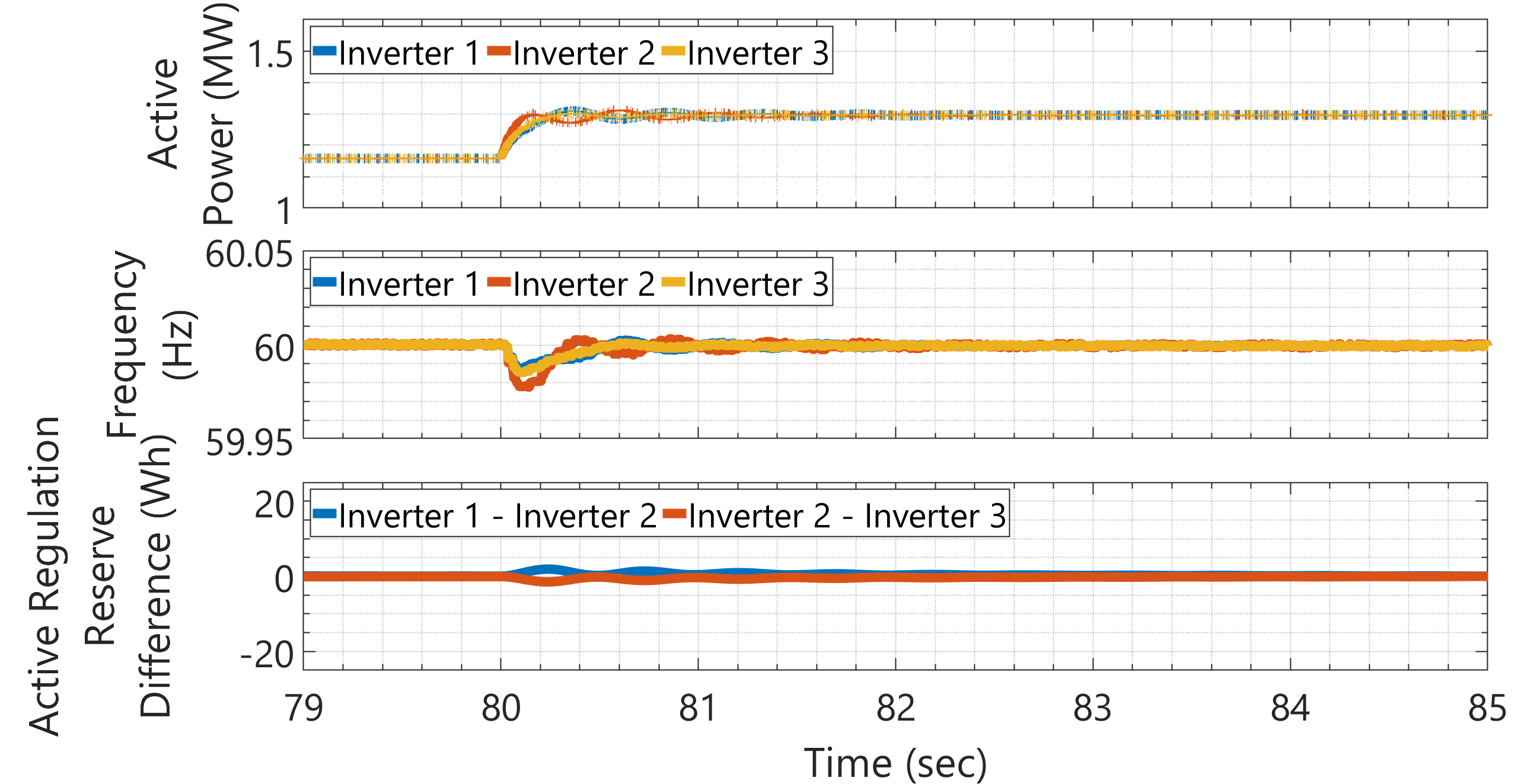}\vspace{-2pt} 
    \caption{%
      From top to bottom: active power, frequency, and active regulation energy reserves using the proposed energy consensus scheme (load increase at 80 seconds for droop-controlled inverters).
    }
    \label{Fig. 9}
  \end{minipage}
\end{figure*}

{\color{black}\subsection{Scenario 1: Droop-controlled Inverters with Active Energy Sharing}}
The first set of simulations involves droop-controlled inverters and utilizes active energy reserve consensus, where $e_{ij}=0.5$ and $f_{ij} = 0$. 10 seconds after the start of the simulation, we perform load pickup with voltage ramp-up to the nominal value $V^*$ over 60 seconds. Fig. \ref{Fig. 6} and Fig. \ref{Fig. 7} show the active power, frequency and active energy reserve difference between inverters for the base DAPI case and the proposed energy consensus scheme respectively. After 80 seconds, a 500 kW increase in load is simulated, as shown in Fig. \ref{Fig. 8} and Fig.  \ref{Fig. 9}. In both cases, it can be seen that regulation energy reserve consensus is achieved while maintaining frequency regulation and power sharing.

\begin{figure*}\vspace{-8pt}
  \begin{minipage}[t]{0.48\textwidth}
    \centering
    \includegraphics[width=0.9\textwidth]{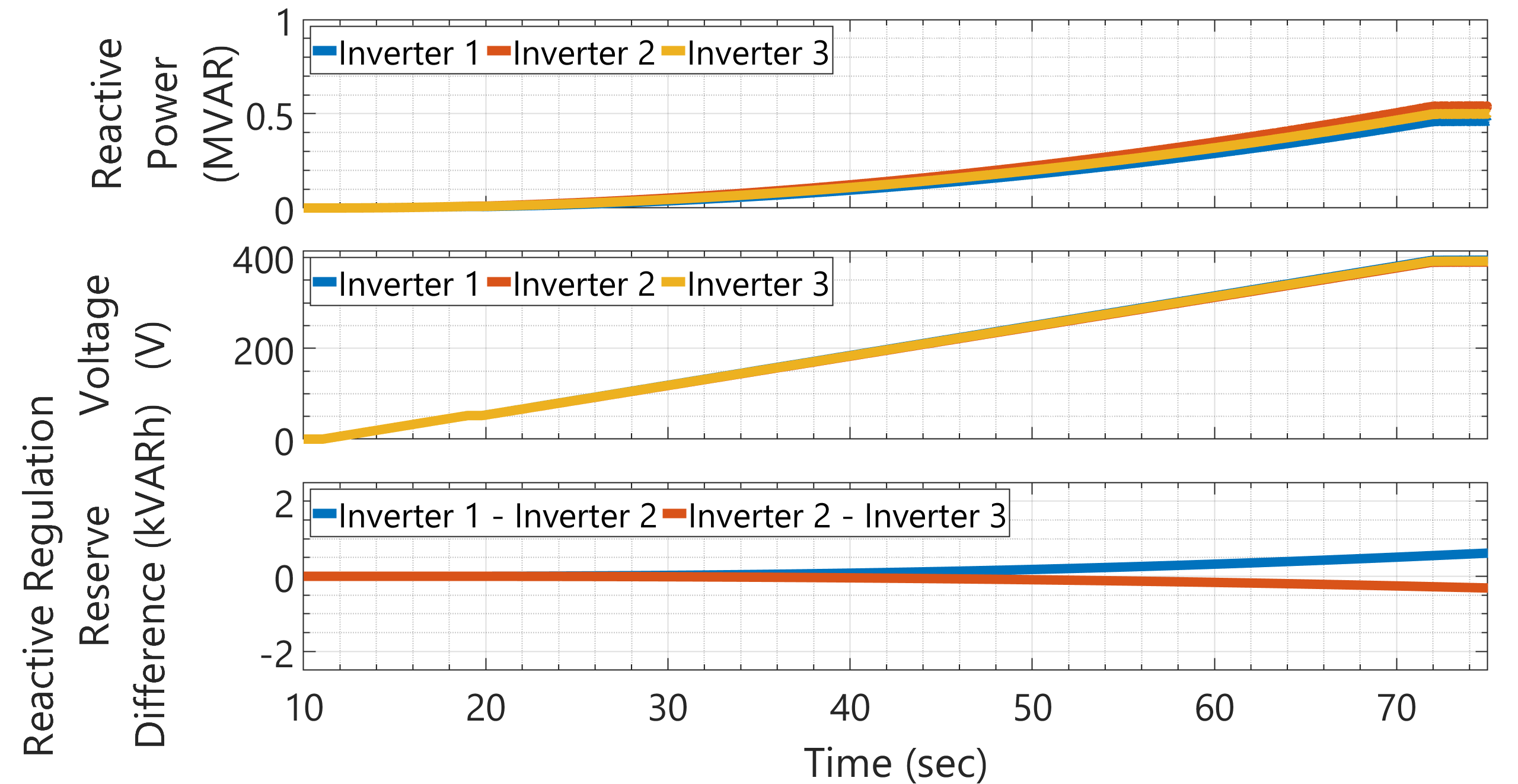}\vspace{-2pt} 
    \caption{%
      From top to bottom: reactive power, voltage, and reactive support reserves using the base secondary control scheme from 10 to 75 seconds (load pickup for VSM-controlled inverters).
    }
    \label{Fig. 10}
  \end{minipage}%
  \hfill
  \begin{minipage}[t]{0.48\textwidth}
    \centering
    \includegraphics[width=0.9\textwidth]{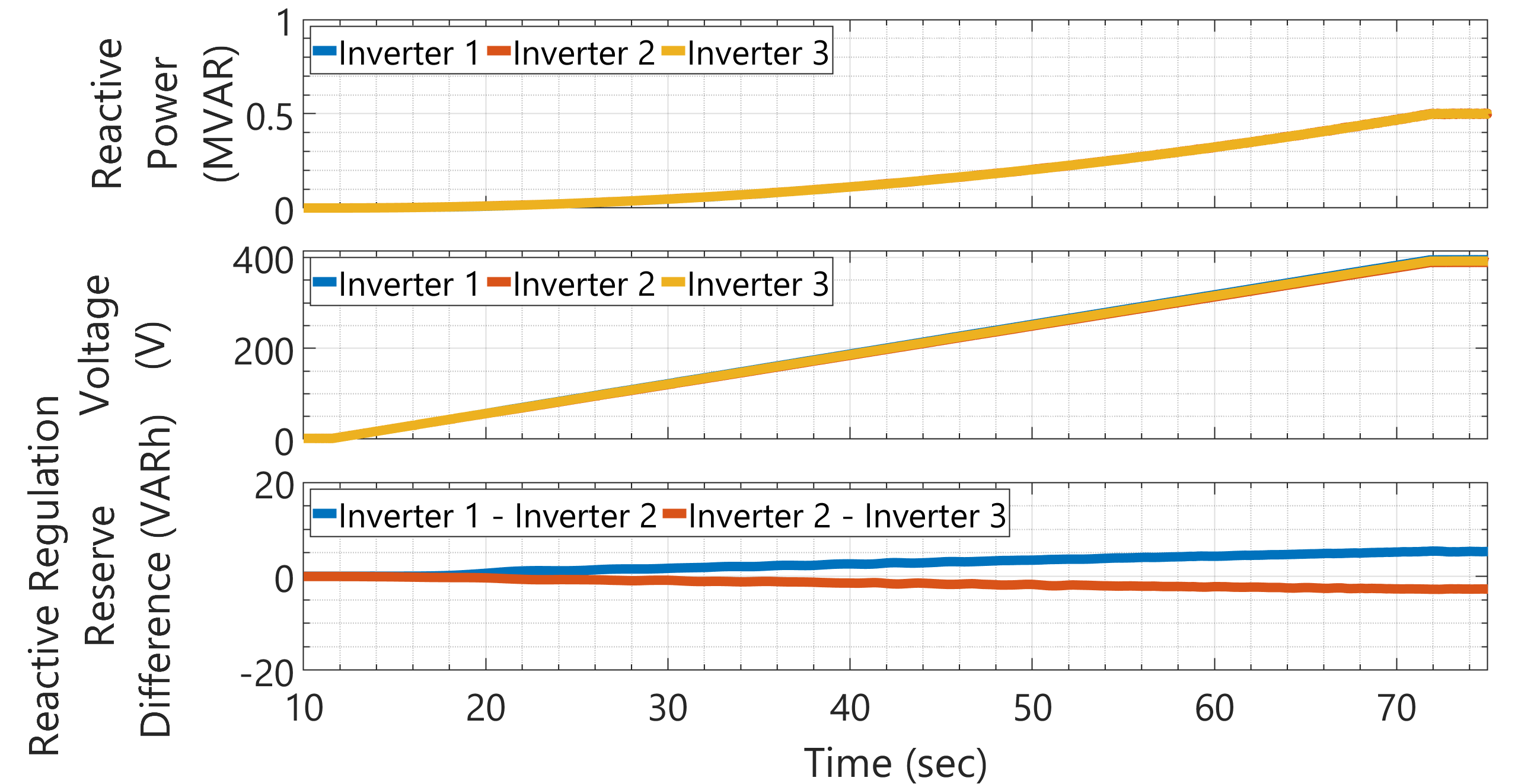}\vspace{-2pt} 
    \caption{%
      From top to bottom: reactive power, voltage, and reactive support reserves using the proposed energy consensus scheme from 10 to 75 seconds (load pickup for VSM-controlled inverters).
    }
    \label{Fig. 11}
  \end{minipage}\vspace{-10pt}
\end{figure*}

\begin{figure*}\vspace{-8pt}
  \centering
  \begin{minipage}[t]{0.48\textwidth}
    \centering
    \includegraphics[width=0.9\textwidth]{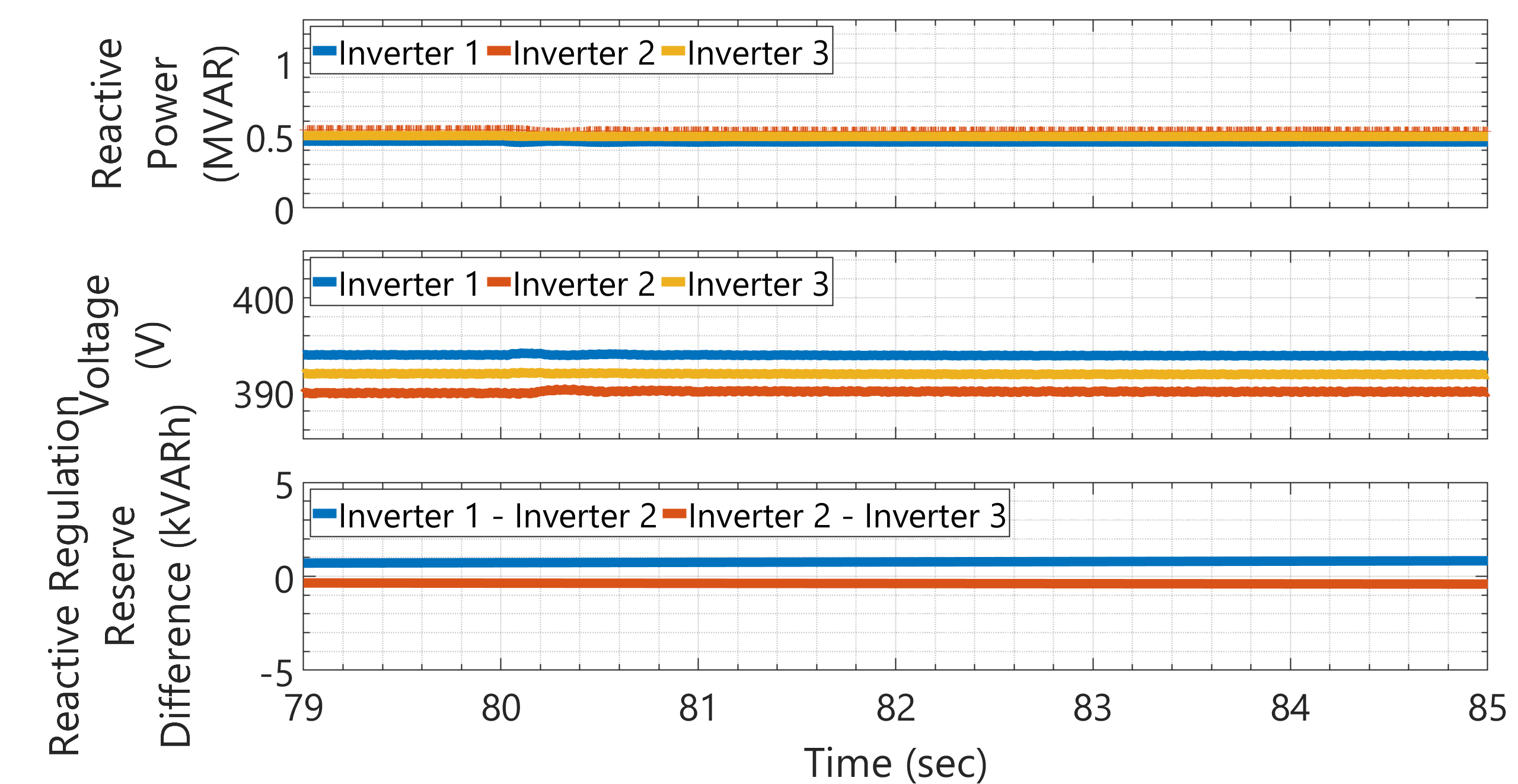}\vspace{-2pt}
    \caption{%
      From top to bottom: reactive power, voltage, and reactive support reserves using the base secondary control scheme (load decrease at 80 seconds for VSM-controlled inverters).
    }
    \label{Fig. 12}
  \end{minipage}%
  \hfill
  \begin{minipage}[t]{0.48\textwidth}
    \centering
    \includegraphics[width=0.9\textwidth]{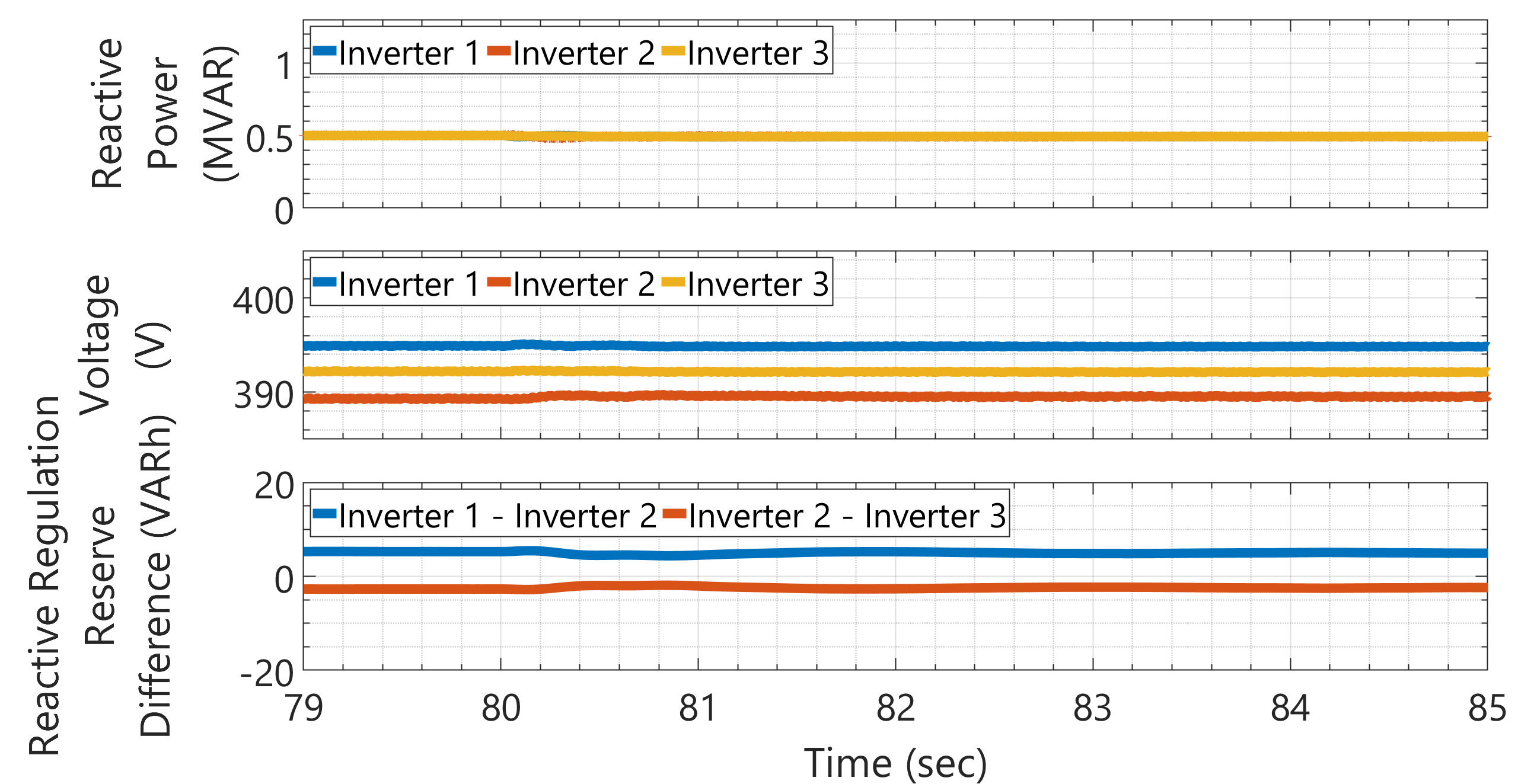}\vspace{-2pt}
    \caption{%
      From top to bottom: reactive power, voltage, and reactive support reserves using the proposed secondary control scheme (load decrease at 80 seconds for VSM-controlled inverters).
    }
    \label{Fig. 13}
  \end{minipage}
\end{figure*}

{\color{black}\subsection{Scenario 2: Virtual Synchronous Machine (VSM)-controlled Inverters with Reactive Support Sharing}}
For the reactive support reserves case, we utilize VSM-based inverters and set $e_{ij}=0$ and $f_{ij}=0.05$. Additionally, a tradeoff between voltage regulation, reactive power sharing and reactive support reserve consensus is achieved. A similar set of scenarios from the previous case is performed, with load pickup in Fig. \ref{Fig. 10} and Fig. \ref{Fig. 11} and load step down in Fig. \ref{Fig. 12} and Fig. \ref{Fig. 13}. The proposed scheme maintains a more constant reactive support reserve compared to the base DAPI scheme, and at a smaller scale. A reserve consensus can be achieved at the cost of removing voltage regulation and reactive power sharing. Nonetheless, an acceptable tradeoff can be achieved that is more beneficial, particularly from a voltage regulation perspective at the distribution level.

\begin{figure*}\vspace{-8pt}
  \begin{minipage}[t]{0.48\textwidth}
    \centering
    \includegraphics[width=0.9\textwidth]{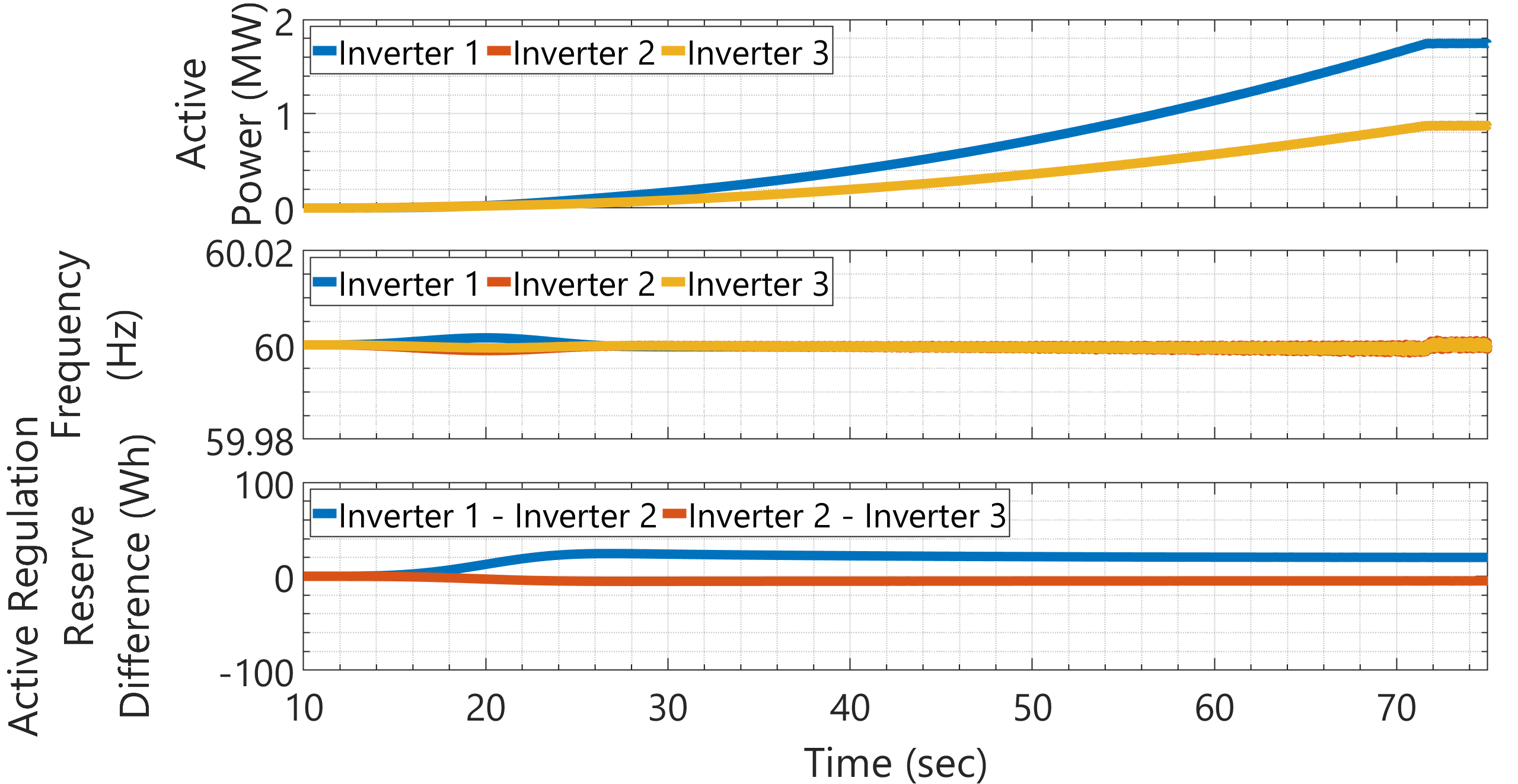}\vspace{-2pt} 
    \caption{%
      From top to bottom: active power, frequency, and active regulation energy reserves using the base secondary control scheme from 10 to 75 seconds (load pickup for droop-controlled and VSM-controlled inverters).
    }
    \label{Fig. 14}
  \end{minipage}%
  \hfill
  \begin{minipage}[t]{0.48\textwidth}
    \centering
    \includegraphics[width=0.9\textwidth]{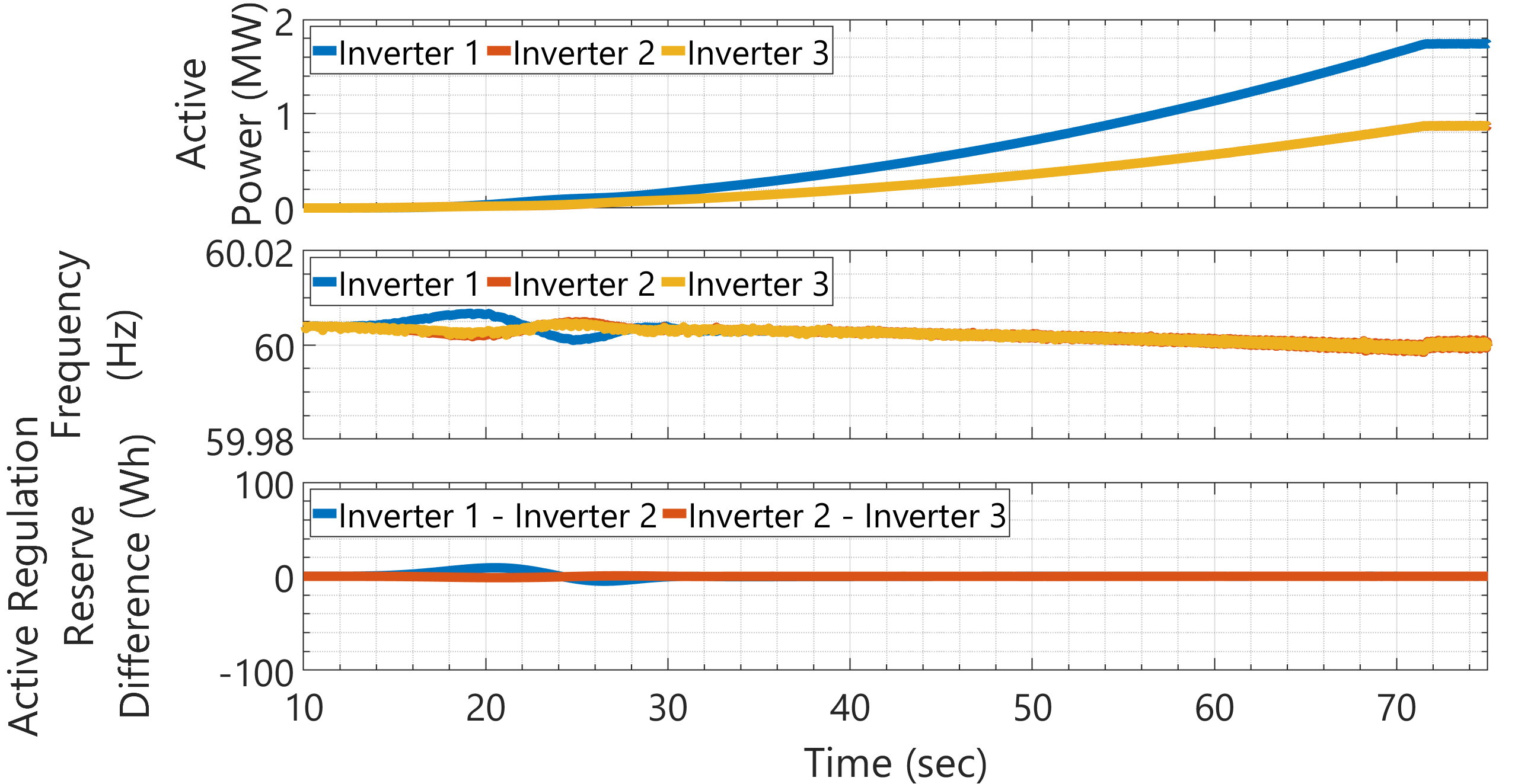}\vspace{-2pt} 
    \caption{%
      From top to bottom: active power, frequency, and active regulation energy reserves using the proposed secondary control scheme from 10 to 75 seconds (load pickup for droop-controlled and VSM-controlled inverters).
    }
    \label{Fig. 15}
  \end{minipage}
\end{figure*}

\begin{figure*}\vspace{-8pt}
  \begin{minipage}[t]{0.48\textwidth}
    \centering
    \includegraphics[width=0.9\textwidth]{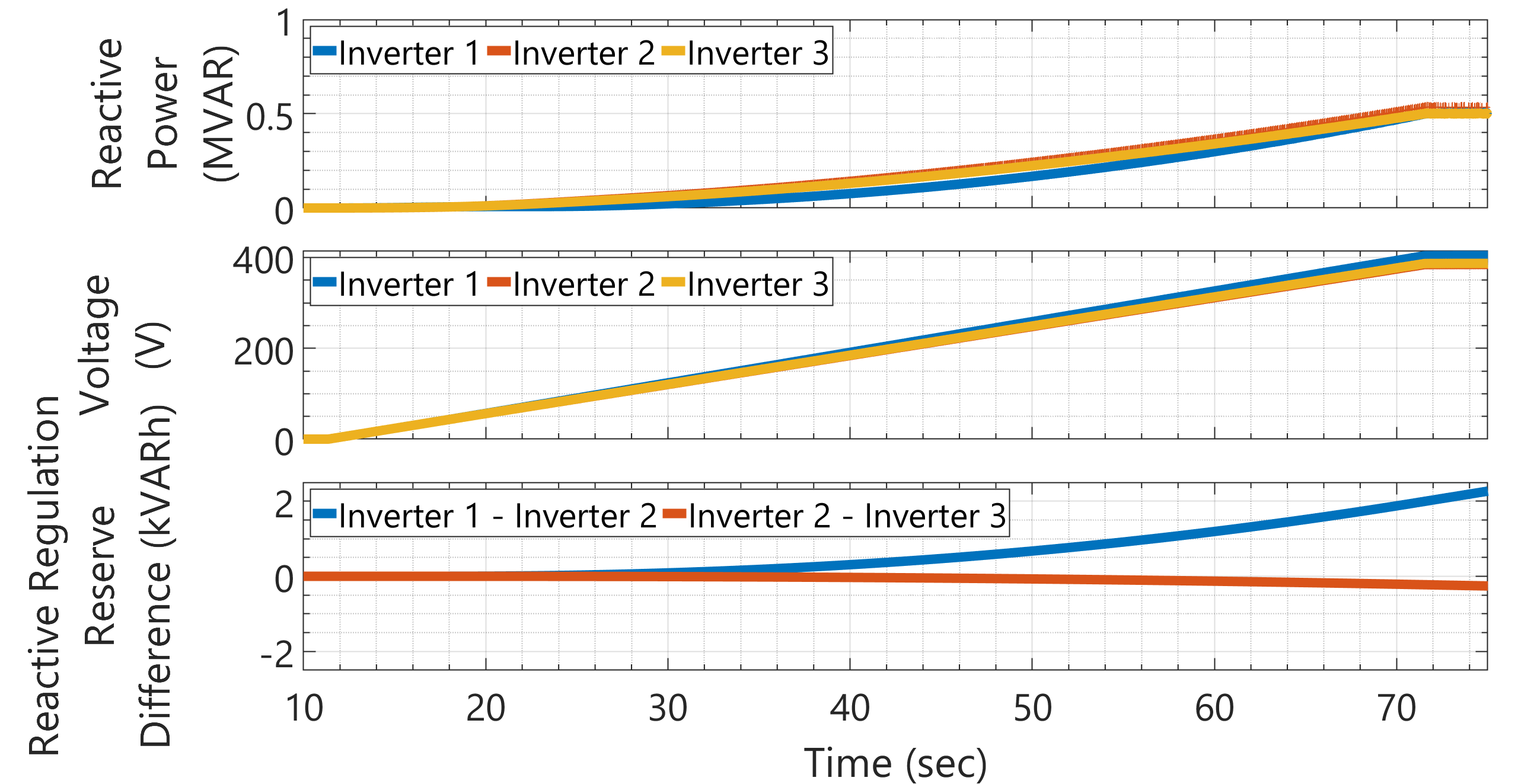}\vspace{-2pt} 
    \caption{%
      From top to bottom: reactive power, voltage, and reactive support reserves using the base secondary control scheme from 10 to 75 seconds (load pickup for droop-controlled and VSM-controlled inverters).
    }
    \label{Fig. 16}
  \end{minipage}%
  \hfill
  \begin{minipage}[t]{0.48\textwidth}
    \centering
    \includegraphics[width=0.9\textwidth]{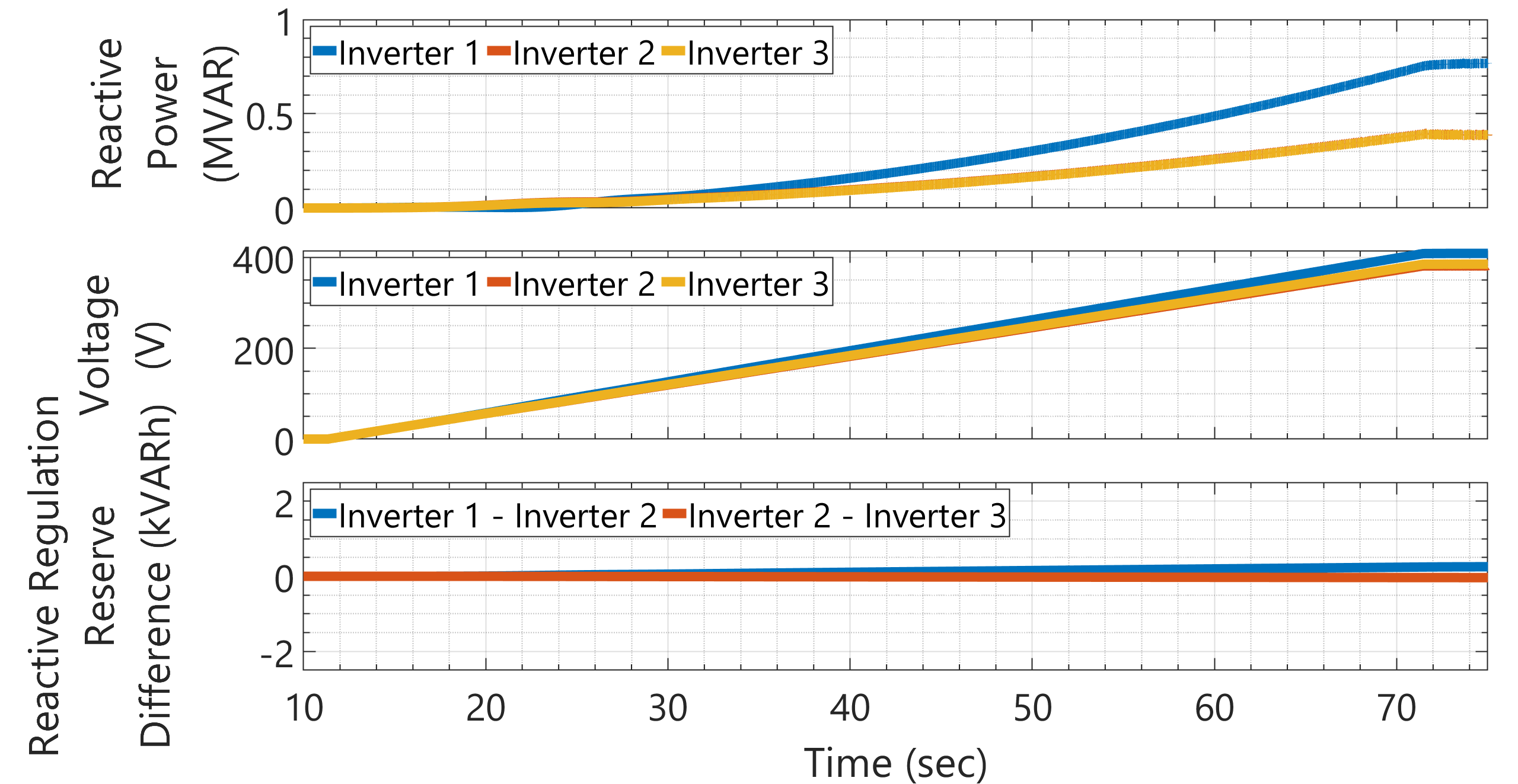}\vspace{-2pt} 
    \caption{%
      From top to bottom: reactive power, voltage, and reactive support reserves using the proposed secondary control scheme from 10 to 75 seconds (load pickup for droop-controlled and VSM-controlled inverters).
    }
    \label{Fig. 17}
  \end{minipage}\vspace{-8pt}
\end{figure*}

\begin{figure*}\vspace{-8pt}
  \begin{minipage}[t]{0.48\textwidth}
    \centering
    \includegraphics[width=0.9\textwidth]{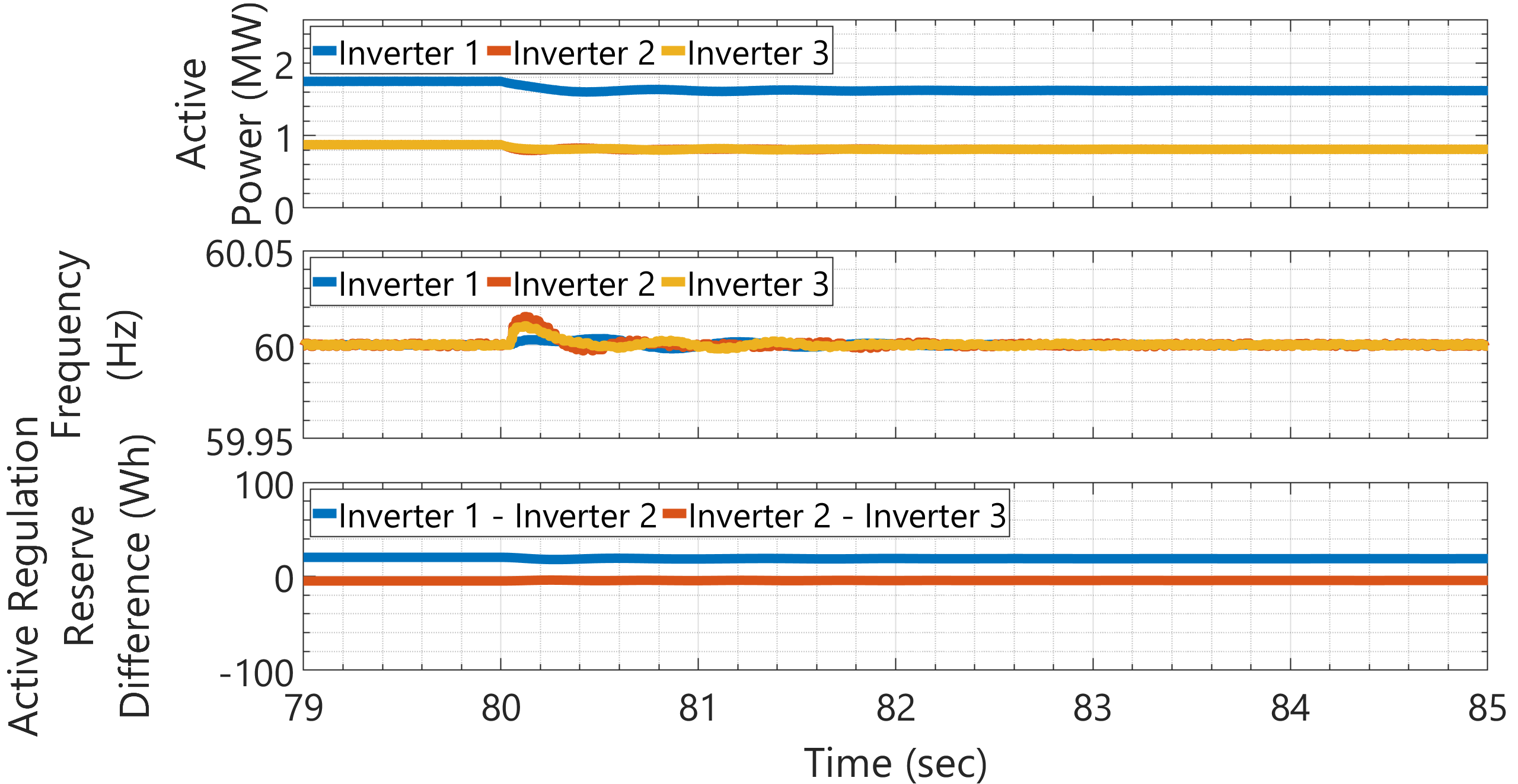}\vspace{-2pt} 
    \caption{%
      From top to bottom: active power, frequency, and active regulation energy reserves using the base secondary control scheme (load decrease at 80 seconds for droop-controlled and VSM-controlled inverters).
    }
    \label{Fig. 18}
  \end{minipage}%
  \hfill
  \begin{minipage}[t]{0.48\textwidth}
    \centering
    \includegraphics[width=0.9\textwidth]{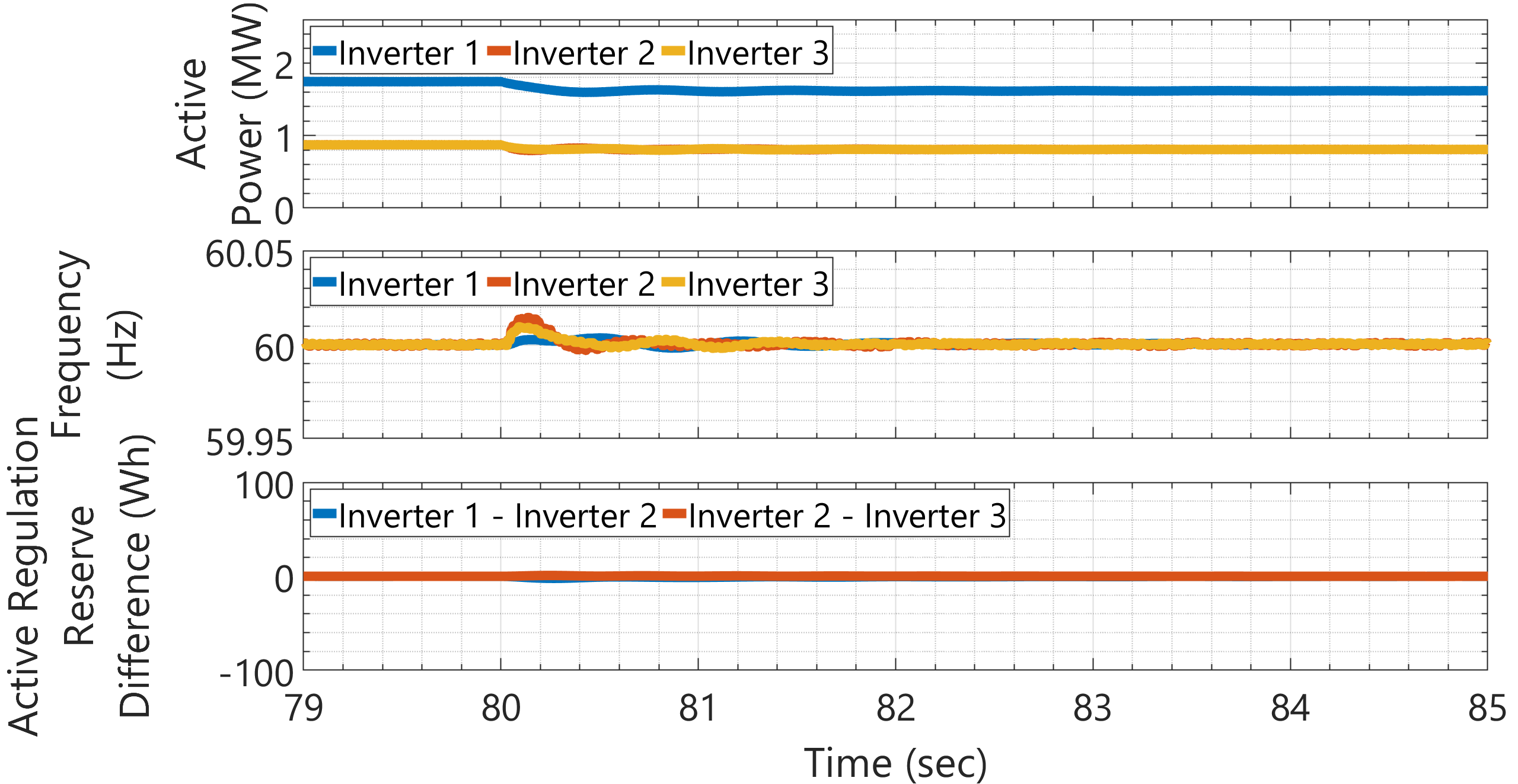}\vspace{-2pt} 
    \caption{%
      From top to bottom: active power, frequency, and active regulation energy reserves using the proposed secondary control scheme (load decrease at 80 seconds for droop-controlled and VSM-controlled inverters).
    }
    \label{Fig. 19}
  \end{minipage}
\end{figure*}

\begin{figure*}\vspace{-8pt}
  \begin{minipage}[t]{0.48\textwidth}
    \centering
    \includegraphics[width=0.9\textwidth]{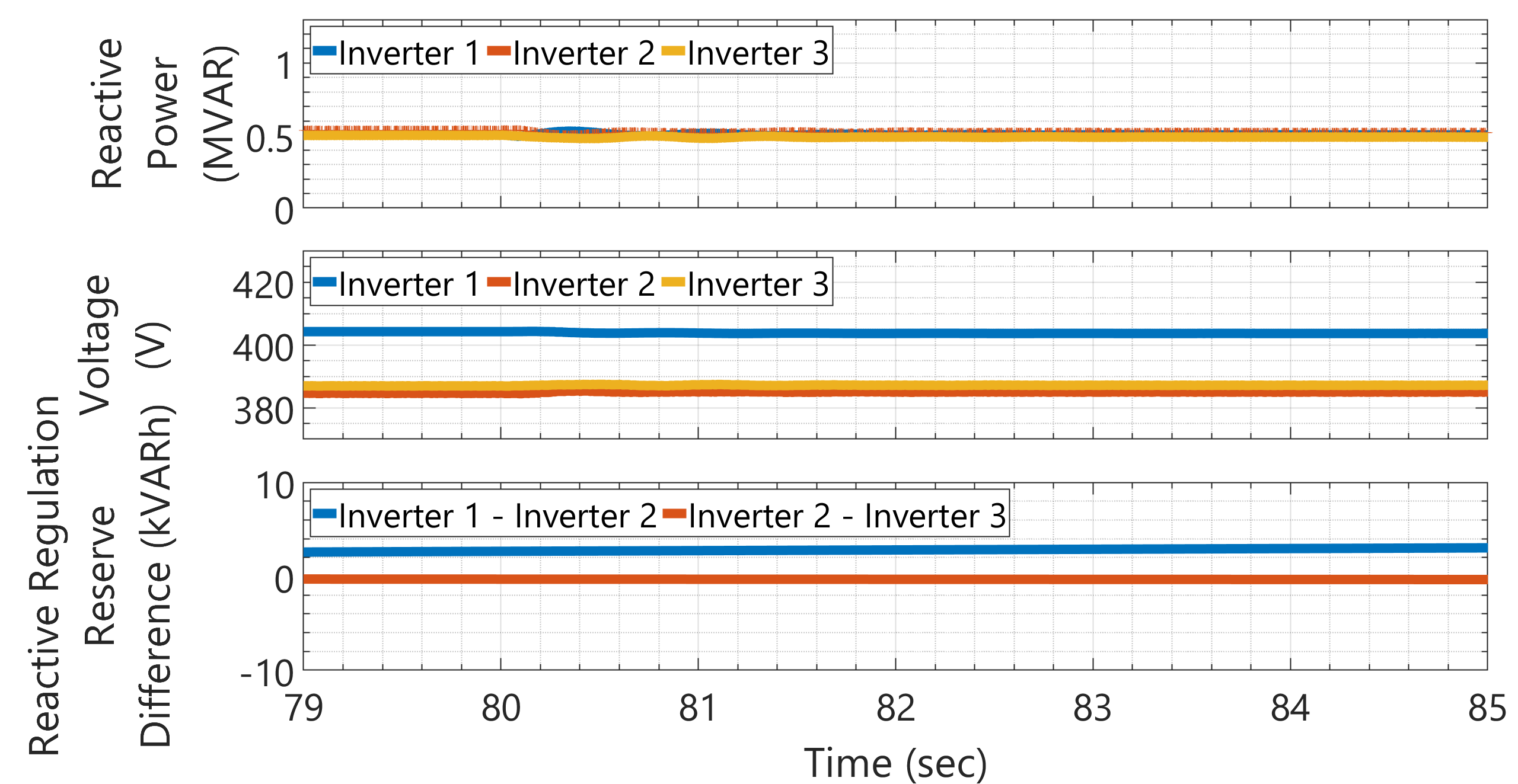}\vspace{-2pt} 
    \caption{%
      From top to bottom: reactive power, voltage, and reactive support reserves using the base secondary control scheme (load decrease at 80 seconds for droop-controlled and VSM-controlled inverters).
    }
    \label{Fig. 20}
  \end{minipage}%
  \hfill
  \begin{minipage}[t]{0.48\textwidth}
    \centering
    \includegraphics[width=0.9\textwidth]{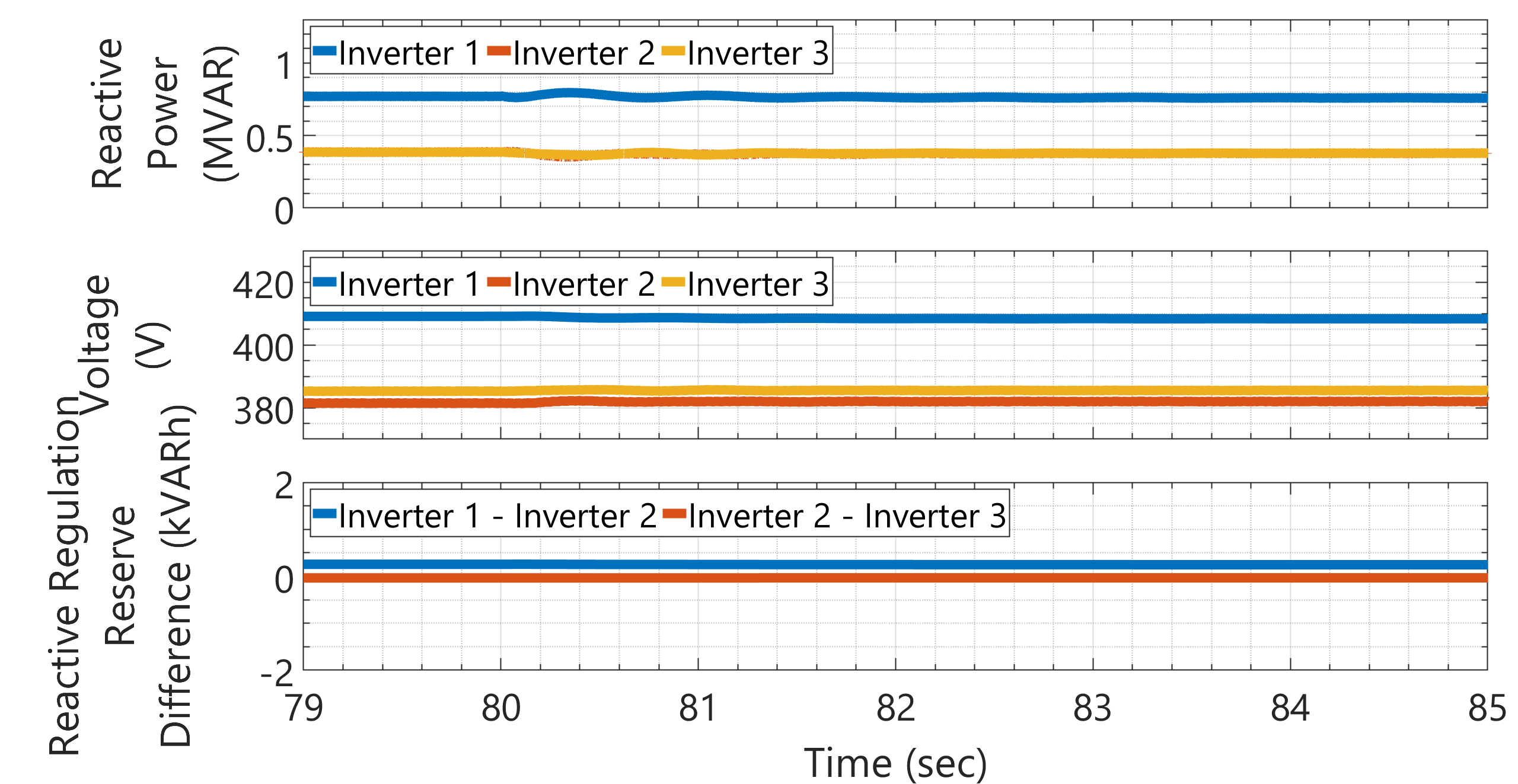}\vspace{-2pt} 
    \caption{%
      From top to bottom: reactive power, voltage, and reactive support reserves using the proposed secondary control scheme (load decrease at 80 seconds for droop-controlled and VSM-controlled inverters).
    }
    \label{Fig. 21}
  \end{minipage}\vspace{-8pt}
\end{figure*}

{\color{black}\subsection{Scenario 3: Heterogeneous Droop and Virtual Synchronous Machine (VSM)-controlled Inverters with Active and Reactive Support Sharing}
Section \ref{Section IV} establishes the stability conditions for homogeneous BESS-based inverters with equivalent capacity and (droop or VSM) dynamics. To illustrate the robustness of the proposed scheme beyond the aforementioned assumptions, we perform CHIL simulations for heterogeneous droop-controlled and VSM-controlled inverters. In this scenario, the droop gains and setpoints are adjusted so that inverter 1 (VSM) is double the capacity of inverter 2 (droop) and inverter 3 (droop). For active and reactive sharing consensus, we set $e_{ij}=0.5$ and $f_{ij}=0.05$. The same set of scenarios from the previous case is performed, with load pickup in Fig. \ref{Fig. 14}, Fig. \ref{Fig. 15}, Fig. \ref{Fig. 16} and Fig. \ref{Fig. 17}, load step down in Fig. \ref{Fig. 18},  Fig. \ref{Fig. 19}, Fig. \ref{Fig. 20} and Fig. \ref{Fig. 21}. It is observed that active power and regulation energy sharing are achieved, along with improved reactive power sharing and regulation energy sharing.}

\subsection{{\color{black}Results Analysis}}
The proposed consensus scheme can achieve equal regulation reserves across inverters. It is interesting to note that with the proposed scheme, energy reserve regulation results in reactive power sharing with minimal compromise to voltage regulation. On the other hand, since the proposed scheme introduces a new integral mode, it can induce adverse oscillations during transient phases. Fig. \ref{Fig. 15} shows slight hunting oscillations during load pickup. Thus, it may require careful tuning, i.e., low energy reserves consensus gains as shown in Table \ref{Table I}.

\vspace{-1pt}
Though it is not the subject of our study, preliminary results indicate that the reserve regulation scheme can be applied to MGs for sharing of active energy and reactive support reserves, which can result in identical SoCs across BESS-based inverters given the appropriate initial conditions. However, such a result may be affected by heterogeneous BESS. In that case, it would depend on the BESS voltage levels, which may require DC/DC voltage conversion to maintain proper SoC balancing. For further validation, more extensive results are necessary for heterogeneous cases in NMGs where inverters are of different capacities and droop gains. Lastly, communication-wise, this scheme is applicable to sparsely linked IBRs. We refer to \cite{b3,b4} for more details. Sparsity can enable NMG level applications, subject to timescale and empirical gain tuning processes.
\vspace{-8pt}
\section{Conclusion}\label{Section VI}
This paper proposed a hierarchical control energy reserves framework that establishes and accounts for inadvertent energy exchange across BESS-based inverters. The proposed framework quantifies service-based reserves \rev{that may inform future} long-term tertiary level headroom. Additionally, we established a DAPI-based energy consensus scheme, to share the active energy and reactive support reserves across inverters. Using this method, we can account for equitable primary level and secondary level regulation energy expenditure from each inverter. Simulations in MATLAB\textsuperscript{\textregistered}/Simulink\textsuperscript{\textregistered} with CHIL validation illustrate the validity of the proposed scheme, achieving frequency and voltage regulation, active and reactive power sharing and sharing of active energy reserves. Future work involves heterogeneous BESS and inverter test cases in NMGs, as well as reserves allocation and adjustment based on tertiary level energy management forecasts.

\manuallabel{Section VII}{VII}\section{Acknowledgment}
This work was funded by the Santa Clara University Latimer Energy Laboratory and the School of Engineering Dean's Excellence Fellowship.

\begin{thebibliography}{00}


\bibitem{b1} M. D. Ilić, “From Hierarchical to Open Access Electric Power Systems," Proceedings of the IEEE, vol. 95, no. 5, pp. 1060-1084, May 2007.

\bibitem{b2} F. Nawaz, E. Pashajavid, Y. Fan and M. Batool, “A Comprehensive Review of the State-of-the-Art of Secondary Control Strategies for Microgrids," IEEE Access, vol. 11, pp. 102444-102459, 2023.
\bibitem{b3} J. W. Simpson-Porco, F. Dörfler and F. Bullo, “Synchronization and power sharing for droop-controlled inverters in islanded microgrids," Automatica, vol. 49, no. 9, pp. 2603-2611, Sep. 2013.
\bibitem{b4} J. W. Simpson-Porco, Q. Shafiee, F. Dörfler, J. C. Vasquez, J. M. Guerrero and F. Bullo, “Secondary Frequency and Voltage Control of Islanded MGs via Distributed Averaging," IEEE Transactions on Industrial Electronics, vol. 62, no. 11, pp. 7025-7038, Nov. 2015.
\bibitem{b5} M. Andreasson, D. V. Dimarogonas, H. Sandberg and K. H. Johansson, “Distributed Control of Networked Dynamical Systems: Static Feedback, Integral Action and Consensus," IEEE Transactions on Automatic Control, vol. 59, no. 7, pp. 1750-1764, July 2014.
\bibitem{b6} J. W. Simpson-Porco, “On Stability of Distributed-Averaging Proportional-Integral Frequency Control in Power Systems," in IEEE Control Systems Letters, vol. 5, no. 2, pp. 677-682, April 2021.
\bibitem{b7} E. Tegling, M. Andreasson, J. W. Simpson-Porco and H. Sandberg, “Improving performance of droop-controlled microgrids through distributed PI-control," 2016 American Control Conference (ACC), Boston, MA, USA, 2016, pp. 2321-2327.
\bibitem{b8} K. Yan, G. Li, R. Zhang, Y. Xu, T. Jiang and X. Li, “Frequency Control and Optimal Operation of Low-Inertia Power Systems With HVDC and Renewable Energy: A Review," IEEE Transactions on Power Systems, vol. 39, no. 2, pp. 4279-4295, March 2024.
\bibitem{b9} Y. Zhang et al., “Investigating Power System Primary and Secondary Reserve Interaction under High Wind Power Penetration," Technical Report, NREL/TP-5D00-64637, Dec. 2016.  
\bibitem{b10} W. Kang, Y. Guan, Y. Yu, B. Arbab-Zavar, J. C. Vasquez and J. M. Guerrero, “Distributed Event-Triggered Secondary Frequency Regulation by Sharing HESS Power in Microgrids," IEEE Transactions on Smart Grid, vol. 15, no. 4, pp. 3375-3389, July 2024.
\bibitem{b11} Y. Chen, J. Zhou, L. Yan, Y. Lin, X. Chen and J. Wen, “Distributed Power Tracking Control of Energy Storage Systems With Aggregated Capacity Estimation," IEEE Transactions on Sustainable Energy, vol. 17, no. 1, pp. 672-683, Jan. 2026.
\bibitem{b12} W. Kang, M. Chen, Y. Guan, L. Tang, J. C. Vasquez and J. M. Guerrero, “Distributed Event-Triggered Optimal Control Method for Heterogeneous Energy Storage Systems in Smart Grid," IEEE Transactions on Sustainable Energy, vol. 13, no. 4, pp. 1944-1956, Oct. 2022.
\bibitem{b13} T. Morstyn, B. Hredzak and V. G. Agelidis, “Distributed Cooperative Control of Microgrid Storage," IEEE Transactions on Power Systems, vol. 30, no. 5, pp. 2780-2789, Sept. 2015.
\bibitem{b14} J. Khazaei and Z. Miao, “Consensus Control for Energy Storage Systems," IEEE Transactions on Smart Grid, vol. 9, no. 4, pp. 3009-3017, July 2018.
\bibitem{b15} D. H. Nguyen and J. Khazaei, “Unified Distributed Control of Battery Storage With Various Primary Control in Power Systems," in IEEE Transactions on Sustainable Energy, vol. 12, no. 4, pp. 2332-2341, Oct. 2021.
\bibitem{b16} L. Zhou, D. Du, M. Fei, K. Li and A. Rakić, “Multiobjective Distributed Secondary Control of Battery Energy Storage Systems in Islanded AC Microgrids," 2021 40th Chinese Control Conference (CCC), Shanghai, China, 2021, pp. 6981-6985.
\bibitem{b17} M. Zare, H. Atrianfar, A. Khorsandi and H. Askarian-Abyaneh, “Distributed Event-Triggered Secondary Control of Islanded Microgrids Incorporating Battery Energy Storage Systems," 2024 28th International Electrical Power Distribution Conference (EPDC), Zanjan, Iran, Islamic Republic of, 2024, pp. 1-7.
\bibitem{b18} \rev{T. Yang and J. Lai, “Brain-Like Distributed Two-Layer SoC Balancing Control of Cyber-Physical Microgrid Clusters," IEEE Transactions on Industrial Cyber-Physical Systems, vol. 3, pp. 567-576, 2025.}
\bibitem{b19} Y. Li, L. Fan, L. Bao and Z. Miao, “CHIL Testbed of Consensus Control-Based Battery Energy Storage Systems," 2020 52nd North American Power Symposium (NAPS), Tempe, AZ, USA, 2021, pp. 1-6.
\bibitem{b20} Z. Afshar, I. Bhogaraju, H. Rahmanei and M. Farasat, “Hierarchical Frequency and SOC Control of Power Grids With Battery Energy Storage Systems," in IEEE Transactions on Power Electronics, vol. 39, no. 7, pp. 7925-7937, July 2024.
\bibitem{b21} S. Som, S. Chakrabarti, S. R. Sahoo, A. Ghosh and X. Liang, “BESS Reserve-Based Frequency Support During Emergency in Islanded Residential Microgrids," in IEEE Transactions on Sustainable Energy, vol. 14, no. 3, pp. 1702-1713, July 2023.
\bibitem{b22} A. Solat, G. B. Gharehpetian, M. S. Naderi, A. Anvari-Moghaddam, “A novel consensus-based control method for SoC balancing of distributed energy storage systems in isolated AC microgrids," International Journal of Electrical Power \& Energy Systems, vol. 176, 2026.
\bibitem{b23} M. Gholami, G. Bianchini and A. Vicino, “Distributed Secondary Control for Battery Energy Storage Systems in AC Microgrids under Multiple Time-Varying Communication Delays," 2024 IEEE 63rd Conference on Decision and Control (CDC), Milan, Italy, 2024, pp. 3483-3488.
\bibitem{b24} \rev{S. Chen, X. Lu and Q. Gong, “Power Optimization for Cyber-Physical Microgrids With Multi-VRB Energy Storage Systems," IEEE Transactions on Industrial Cyber-Physical Systems, vol. 4, pp. 269-279, 2026.}
\bibitem{b25} T. A. Fagundes et al., “Battery Energy Storage Systems in Microgrids: A Review of SoC Balancing and Perspectives," IEEE Open Journal of the Industrial Electronics Society, vol. 5, pp. 961-992, 2024.
\bibitem{b26} Y. Rebours and D. Kirschen, “What is spinning reserve?," The University of Manchester, 2005.
\bibitem{b27} California Independent System Operator, “Special Report on Battery Storage," Department of Market Monitoring, 2023.
\bibitem{b28} National Energy System Operator, “Response Stacking Guidance," 2025.
\bibitem{b29} S. Mohammad Tayyeb and R. Krishan, “Data-Centric Decentralised Controller for Effective P-F and Q-V Control in AC Microgrids," 2023 IEEE 3rd International Conference on Sustainable Energy and Future Electric Transportation (SEFET), Bhubaneswar, India, 2023, pp. 1-6.
\bibitem{b30} J. Santos Döhler, “Hierarchical Control and Inverter Functionalities in Microgrids. Modeling, Simulation, and Case Studies," Digital Comprehensive Summaries of Uppsala Dissertations from the Faculty of Science and Technology, 2603, Uppsala: Acta Universitatis Upsaliensis, Nov. 2025.
\bibitem{b31} X. Chen, S. Bu and I. Kocar, “Grid-Forming IBRs Under Unbalanced Grid Conditions: Challenges, Solutions, and Prospects," in IEEE Transactions on Sustainable Energy, vol. 16, no. 4, pp. 3031-3047, Oct. 2025.
\bibitem{b32} B. Johnson et al., “A Generic Primary-control Model for Grid-forming Inverters: Towards Interoperable Operation \& Control," 2022 55th Hawaii International Conference on System Sciences (HICSS), pp. 3398–3407, 2022.
\bibitem{b33} J. Zhang, Y. Men, L. Ding and X. Lu, “Secondary Frequency and Voltage Regulation for Inverter-Based Microgrids: A Sparsity-Promoting DAPI Control Approach," IEEE Transactions on Control Systems Technology, vol. 32, no. 4, pp. 1512-1519, July 2024.
\bibitem{b34} Pacific Gas and Electric Company, “Electric Rule No. 1: Definitions," Pacific Gas and Electric Tariff Book, 2007.
\bibitem{ancillarypaper} Y. G. Rebours, D. S. Kirschen, M. Trotignon and S. Rossignol, "A Survey of Frequency and Voltage Control Ancillary Services—Part II: Economic Features," in IEEE Transactions on Power Systems, vol. 22, no. 1, pp. 358-366, Feb. 2007.
\bibitem{b35} R. Olfati-Saber, J. A. Fax and R. M. Murray, “Consensus and Cooperation in Networked Multi-Agent Systems," in Proceedings of the IEEE, vol. 95, no. 1, pp. 215-233, Jan. 2007.
\bibitem{b36} L. Xiao, S. Boyd and S.-J. Kim, “Distributed average consensus with least-mean-square deviation," Journal of Parallel and Distributed Computing, vol. 67, no. 1, pp. 33–46, Jan. 2007.




\end{thebibliography}
\end{document}